\documentclass[aps,pre,onecolumn,nofootinbib,longbibliography]{revtex4-1}
\usepackage{times}
\usepackage{graphicx}
\usepackage{amsmath}
\usepackage{amsfonts}
\usepackage{amssymb}
\usepackage{graphicx}
\usepackage{color}
\begin{document}
\title{Dynamics of Rod like Particles in Supercooled Liquids - Probing 
Dynamic Heterogeneity and Amorphous Order}
\author{Anoop Mutneja}
\email{anoopm@tifrh.res.in}
\author{Smarajit Karmakar}
\email{smarajit@tifrh.res.in}
\affiliation{
Tata Institute of Fundamental Research, 
36/P, Gopanpally Village, Serilingampally Mandal,Ranga Reddy District, 
Hyderabad, 500107, Telangana, India } 

\begin{abstract}
Probing dynamic and static correlation in glass-forming supercooled liquids
has been a challenge for decades in spite of extensive research. 
Dynamic correlation which manifests itself as Dynamic Heterogeneity is 
ubiquitous in a vast variety of systems starting from molecular glass-forming
liquids, dense colloidal systems to collections of cells. On the other hand 
mere concept of static correlation in these dense 
disordered systems remain somewhat elusive and its existence is still actively 
debated. We propose a novel method to extract both dynamic and static 
correlations using rod like particles as probe. This method can be implemented
in molecular glass-forming liquids in experiments as well as in
other soft matter systems including biologically relevant systems. 
We also rationalize the observed log-normal like distribution of rotational 
decorrelation time of elongated probe molecules in reported experimental 
studies along with a proposal of a novel methodology to extract 
dynamic and static correlation lengths in experiments.
\end{abstract}
\maketitle
\section*{Introduction}
Being structurally disordered, constituent particles or molecules in both 
the liquid and glass phases, experience 
variable local environments. This can be easily ignored for high temperature 
liquids but for supercooled liquids it manifests itself in spatial distribution 
of particle's mobility, from nearly stuck to fairly moving. This in turn 
gives rise to complex behaviour in their bulk properties like viscosity, 
structural relaxation time as well as diffusion constant. 
All of these complex dynamical behaviour is termed in literature as  
Dynamical Heterogeneity (DH)\cite{HeteroRev,Berthier2011}. It has been shown that 
the slow and fast moving local regions form clusters
\cite{Keys,Kawasaki_Tanaka,TanakaMRCO} which 
lead to strong spatial variation in local relaxation times of the system. 
These spatial variation in local relaxation times adds up to an overall 
non-exponential relaxation, observed in various 
experimental \cite{Richert_2002} and simulation \cite{Perera1999} studies.
Non-Gaussian behaviour with universal exponential tail \cite{pinakivanhove} 
in the distribution of particles' displacements or the van Hove function 
(See SI for definition) is also a direct manifestation of the same dynamical 
heterogeneity. 

On the other hand, in the supercooled regime, the 
relaxation time or the viscosity of the system also increases very 
drastically with little decrease (increase) in temperature (density). 
In \cite{KDSPNAS2009, KDSAnnualReview, KDSROPP}, it was argued that 
the growing dynamical heterogeneity length scale is not causally 
related to the growth of the timescale or viscosity. Thus existence of 
yet another length scale became necessary to rationalize the rapid 
growth of relaxation time and in \cite{KDSPNAS2009} it was proposed 
that a static length 
scale also grows with decreasing temperature. It is important to 
note that existence of such a static length scale is consistent with 
the predictions of Random First Order Transition (RFOT) Theory 
\cite{RFOT1,RFOT2}. In \cite{PTS}, a new correlation function, 
known as Point-to-Set (PTS) 
correlation function (see SI for definition), was proposed and estimated 
in various model glass-forming liquids to extract the static length 
scale. Growth of the PTS length scale is found to be connected with 
growth of the relaxation time with supercooling. Various other 
measures of similar static length scale are also
found to be consistent with each other \cite{SmarajitPRL2013,SKPhysA}. The 
underlying structural order related to PTS length scale is often 
referred to as ``Amorphous Order''. It is now well 
established that there are indeed two different length 
scales that grow while approaching glass transition \cite{Kob2011,Tah2018PRL}, 
although a possible mutual relation between these two length scales remain 
poorly understood\cite{RajsekharJCP2017}. It is important highlight that
there are lot of effort of identify structural motifs that can be related
to the growth of static length scale \cite{TanakaNatureReview,Tanaka_Paddy}       

To quantify dynamic heterogeneity, one often measures 
the length scales and time scales of different mobility clusters and 
their variation while approaching glass transition. The dynamical 
heterogeneity length scale, $\xi_D$ is computed in general
using \cite{Book1,Dasgupta1991} the peak value of  fluctuations of  the total
mobility characterized by four-point susceptibility, $\chi_4(t)$
(see SI for definition) which is often assumed to be related to 
$\xi_D$ as $\chi_4^P \sim \xi_D^{2-\eta}$ with $\eta$ being an unknown 
exponent. $\xi_D$ can also be computed from the spatial correlation in 
the particles' mobility field\cite{PhysRevLett.82.5064,POOLE199851,Tah2019}. 
Recently, in Ref.~\cite{Bhowmik2018,Block}, 
the non-Gaussian nature of the van Hove function is used to probe 
the dynamical heterogeneity length scale very efficiently by 
systematically coarse-graining the system at varying length scales. 
The idea is that upon coarse-graining over the length scale 
comparable or larger than dynamical heterogeneity length scale,
one would expect that the distribution of particle's displacement 
or the van Hove function will tend to become Gaussian. In this work 
we have studied dynamics of rod-like particles in supercooled 
liquids with varying length of the rods to similarly probe %the SK 
response of the system at varying coarse-graining length scales 
and extract the dynamical heterogeneity length scale. Thus, this 
method might become more accessible to experiments for measuring the 
dynamic heterogeneity length scale in various molecular glass-forming 
liquids as well as in colloidal glasses.

%Direct experimental measure of $\chi_4$ for molecular liquids is difficult as one needs to spatially and temporally resolve the trajectories of all the constituents particles in the system in order to directly of all constituents particles in the system in order to directly compute $\chi_4$ or the displacement-displacement correlation function. %SK
Direct experimental measure of $\chi_4$ or 
displacement-displacement correlation function for molecular liquids 
is not possible as one needs to spatially and temporally resolve the trajectories 
of all constituents particles in a system. 
Although similar measurement can be done for colloidal or granular systems 
\cite{Weeks2000,Kegel2000,ColloidVijay,Bonn2003,Reentrant}. Thus for molecular liquids, one often measures $\chi_4(t)$ 
indirectly as shown in Ref.~\cite{Berthier2005}. It was shown that 
a suitable dynamical response function $\chi_x(t)$ to an induced perturbation 
variable,`$x$', {\it e.g.} density fluctuations in colloidal glasses 
or temperature fluctuation in molecular glass-forming liquids will be 
related to $\chi_4(t)$. One thus estimates $\chi_4(t)$ 
by using linear response formalism and fluctuation theory, but an 
accurate estimate of the length scale will still not be possible
as the exponent, $\eta$ is apriori unknown \cite{JCPBB, 
JCPBB1,PhysRevLett.97.195701}. Similarly, experimental 
measure of growing amorphous order is also very limited and a direct 
evidence of such a growing static length scale came from the measurement
of fifth-order dielectric susceptibility ($\chi_5(t)$) in supercooled 
glycerol and propylene \cite{Albert2016} and via random pinning using 
holographic optical tweezers \cite{AKSoodRGanpati}. The intricacy of the 
experimental measurement immediately tells us that an accurate measurement of 
growing amorphous order is still very hard. Thus it is not very surprising that 
we do not have strong experimental evidence of growth of both dynamical 
heterogeneity length scale as well as the static length scale of amorphous 
order in molecular glass-forming liquids. An experimentally realizable 
proposal for possible measurements of these two important length scales 
will definitely be of importance for understanding the puzzle of glass 
transition.  

In particular, the experiments on the rotational dynamics of a single probe 
in the form of dye molecules\cite{Zondervan2007} and nano-rods \cite{Yuan2013} 
are very encouraging. In \cite{Yuan2013}, the rotational correlation time 
($\tau$) of  gold nano-rods in supercooled glycerol are measured. This time 
scale is found to increase with decreasing temperature. The distribution of 
$\tau$ is surprisingly found to be log-normal in nature whose variance
increases with decreasing temperature, indicating a possible 
increase in dynamic heterogeneity. Similar results were also obtained for 
single dye molecule experiments in supercooled glycerol \cite{Zondervan2007}.
Note that nano-rods are of much larger in size than usual dye molecules. 
The appearance of log-normal distribution in rotation time 
itself conveys the finite probability of rod or molecule to be rotationally 
immobile or just vibrating for most of the time. This also implies 
the existence of heterogeneity at both the length scales of gold nano-rod 
and of single dye molecule. Although existence of dynamical heterogeneity 
is evident from these experimental measurements, a direct measure of dynamical 
heterogeneity length scale is still not available. In this work, we propose 
that growth of both dynamical and static length scales can be obtained 
using similar single molecule probe experiments by systematically varying 
the length of probe molecules and studying their rotational relaxation dynamics.

Our proposed methodology is simply to study the rotational dynamics of 
rod-like probe molecules as done experimentally in 
\cite{Yuan2013,Zondervan2007} but look at the changes in dynamics as 
one varies the length of probe rod as schematically shown in the top 
left panel of Fig:\ref{NNP}. To quantify the heterogeneity in dynamics, 
we use non-normal parameter for rotational diffusion of the rods. This 
is similar to the Binder cumulant or the non-Gaussian parameter usually
studied in the context of studying dynamic heterogeneity in systems with
spherical particles. In Ref.~\cite{Jain2017}, it has been analytically 
shown that the following will be the appropriate non-normal parameters for distributions 
$P_{2D}(\phi,t)$ and $P_{3D}(\theta,t)$, where $\phi(t)\in[-\pi,\pi]$ 
is the polar angle in 2D system and $\theta(t)\in[0,\pi]$ is the azimuthal 
angle in 3D system (assuming the rods are initially placed along the 
positive x-axis in 2D ($\phi=0$) and along positive z-axis ($\theta=0$) 
in 3D).
\begin{equation}
%\begin{split}
\alpha_{rot,2D} = \frac{1}{3}\frac{\left\langle|\hat{u}(t)-\hat{u}_0|^4\right\rangle}{\left(\left\langle|\hat{u}(t)-\hat{u}_0|^2\right\rangle\right)^2}
-\frac{1}{24}\left\langle|\hat{u}(t)-\hat{u}_0|^2\right\rangle\\
 \times\left( \left\langle | \hat{u}(t)-\hat{u}_0|^2\right\rangle - 8 \right) -1
%\end{split}
\end{equation}
\begin{equation}
\alpha_{rot,3D} = \frac{1}{2}\frac{\left\langle|\hat{u}(t)-\hat{u}_0|^4\right\rangle}{\left(\left\langle|\hat{u}(t)-\hat{u}_0|
^2\right\rangle\right)^2}
+\frac{1}{6}\left\langle|\hat{u}(t)-\hat{u}_0|^2\right\rangle\\ -1 .
\end{equation}
$\hat{u}(t)$ is  the orientation unit vector of rod at time `$t$' and 
$\hat{u}_0$ is the orientation vector at time $t=0$, implying 
$\theta = cos^{-1}(\hat{u}(t).\hat{u}_0)$ in 3D and 
$\phi = cos^{-1}(\hat{u}(t).\hat{u}_0)$ in 2D. These non-normal  
parameters of the distributions $P_{2D}(\phi,t)$ and 
$P_{3D}(\theta,t)$ will go to zero if there is no heterogeneity 
present in the system. Thus value of this non-normal parameter would 
then quantify the variation in rotational diffusion constant of the 
rod and hence the dynamic heterogeneity.	

If the liquid is at high temperature (or low density) then one expects 
the non-normal parameters, $\alpha_{rot,2D}(t,T)$ or 
$\alpha_{rot,3D}(t,T)$ to remain zero for all rod lengths and at all times, 
but for supercooled liquid in the presence of dynamic heterogeneity, 
the non-normal parameters will be non-zero and one can expect to see a maximum 
at time scale around $t = \tau_\alpha$ similar to $\chi_4(t,T)$. One also 
expects that peak value of the non-normal parameters ($\alpha^P_{rot,2D}$, $
\alpha^P_{rot,3D}$) should grow with decreasing temperature consistent
with the growth of heterogeneity in the system. It is somewhat intuitive 
to understand that the values of $\alpha^P_{rot,2D}$, $\alpha^P_{rot,3D}$ 
too will decrease with increasing rod length as the rod will now
experience the collective dynamical response of the surrounding liquid 
medium averaged over a volume of linear size comparable to the length of 
rod. Thus in principle, we will have the measure of the heterogeneity at 
various length scales using which the calculation of dynamic heterogeneity 
length scale should not be very difficult. Also, as Finite-Size-Scaling (FSS) 
of structural relaxation time $\tau_\alpha$ \cite{KDSPNAS2009} gives us 
static length scale we can be hopeful to be able to obtain the same from the 
rotational correlation time of the rod and its distributions along with a 
possible understanding of the experimentally observed log-normal distribution 
of rotational correlation time ~\cite{Yuan2013,Zondervan2007} of the 
probe molecules.

In this work, we have done extensive 
simulation of three model glass-forming liquids as discussed in detailed 
in the Method section. These three models are referred in the rest of 
the article as {\bf 3dKA}, {\bf 3dHP} and {\bf 2dmKA} models. The 
details of the parameters of the models and the techniques used in 
performing Molecular Dynamics simulations in the presence of the rod-like
particles can be found in the Method section. The rest of the paper can 
be broadly separated into two parts. In the first part, we discuss in 
detail the scaling analysis performed to extract the 
dynamical heterogeneity length scales from the rod length dependence of 
rotational non-normal parameter both in two and three dimensional systems. 
Then in the second part, we discuss First Passage time (FPT) distribution 
of the rod molecules 
in the liquid with increasing supercooling and how one can extract the 
static length scale from that.

\section*{Results}
\subsection*{Non-Normal Parameters and Dynamic Heterogeneity Length}
\begin{figure*}[!ht]	
\includegraphics[width=1.0\textwidth]{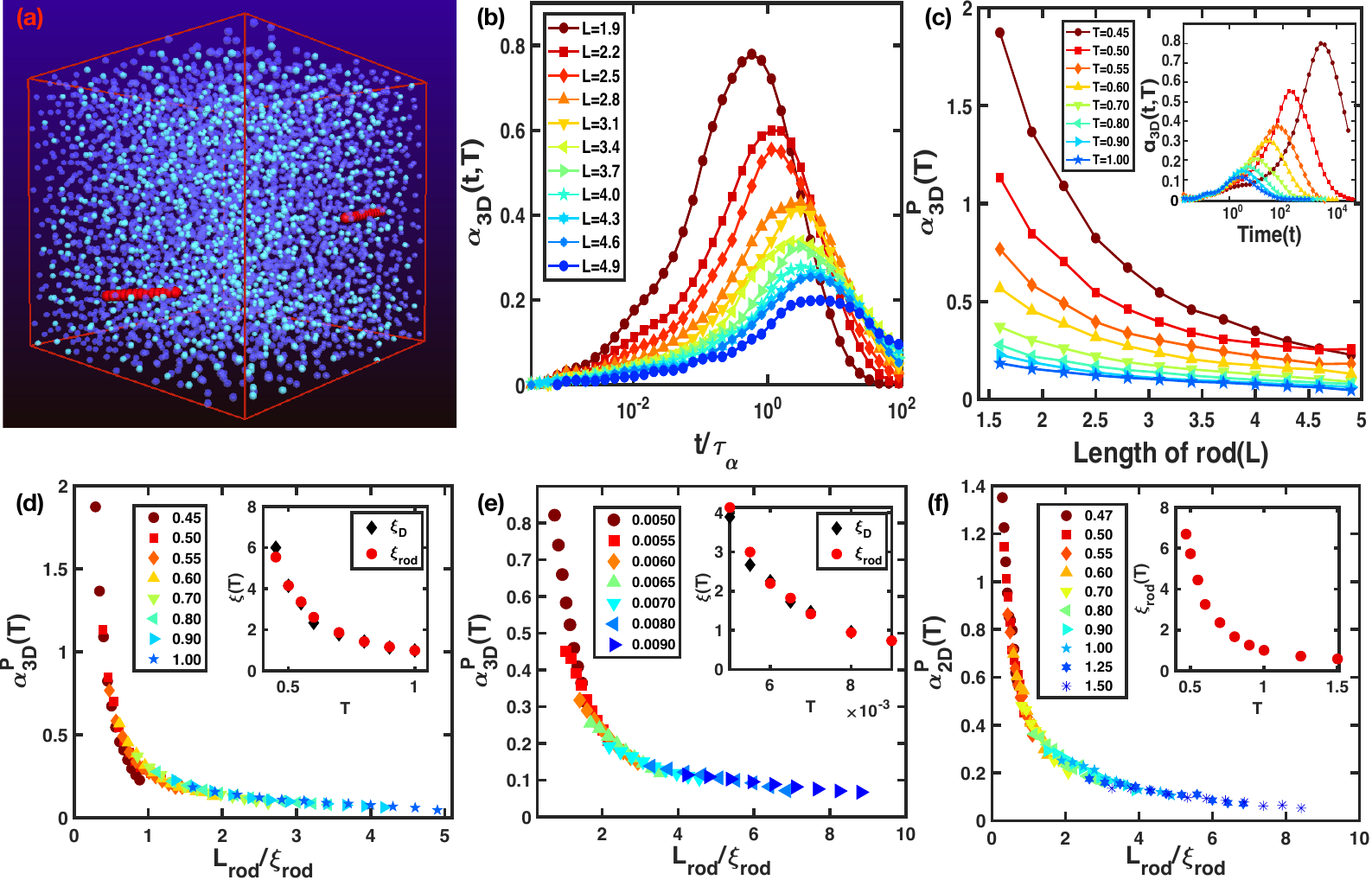}
	\caption{(a) Schematic representation of rod particles in the 
	supercooled liquid medium. (b) Evolution of Non-Normal parameter  
	($\alpha_{rot,3D}(t)$) for 
	different size of the rod in 3dKA supercooled liquid model at 
	temperature $T=0.50$. Inset of (c) Evolution 
	of Non-Normal parameter ($\alpha_{rot,3D}(t)$) for rods of length
	$L_{rod} = 2.5$ in the same supercooled liquid model at different 
	temperatures. 
	(c) Variation in peak value of Non-Normal parameter  $\alpha^P_{3D,rot}$ 
	with length of the rod in 3dKA system at different 
	supercooling temperatures. Bottom panels: Collapse of non-normal 
	parameter obtained by scaling the rod length with appropriate correlation 
	length scale $\xi_{rod}$ for all three systems (3dKA, 3dHP and 2dMKA from 
	(d) to (f)). In the insets of these plots, the scaling length scale 
	is compared with the dynamic length scale of the parent liquid obtained 
	by other conventional methods~\cite{Block} (see text for details). 
	Both the length scales are found to be in good agreement with each other.}
	\label{NNP}
%\label{scalling_collapse}
%\end{figure*}
\end{figure*}

Fig.~\ref{NNP}(b) shows the time evolution of non-normal parameter
$\alpha_{rot,3D}$ for 3dKA  model at $T = 0.50$. Clearly it shows a maximum
at $t\sim\tau_\alpha$ indicating that such analysis indeed picks up the 
heterogeneity of the parent liquid. It can also be seen that the value 
of peak goes down with increasing rod length which validates the correctness 
of our assumption that the dynamically heterogeneous environment experienced
by the rod particles gets averaged out with increasing length of the rod. 
In the inset of Fig: \ref{NNP}(c) we show similar data for a given rod
with decreasing temperature. This results show the increase of heterogeneity 
in parent liquid at the probing length scale of the order of the rod length
with decreasing temperature. Fig: \ref{NNP}(c) shows the variation of 
$\alpha_{rot,3D}^P$ as a function of rod length, for different temperatures 
for 3dKA model. Results are very similar for other models as shown in SI. 

The dynamic heterogeneity as probed by the rod decreases with increasing rod 
length, one can expect that the behaviour changes when the rod length is 
comparable to the dynamic heterogeneity length scale, $\xi_D$ at the studied
temperature. Thus one can attempt to do a scaling plot of the non-normal
parameters for different lengths of the rod and temperatures using the 
dynamical heterogeneity length scale. If the physics is governed by 
the dynamical heterogeneity length scale then one would expect a master 
plot when $\alpha_{rot,3D}^P$ or $\alpha_{rot,2D}^P$ is plotted as a function
of $L_{rod}/\xi_{D}(T)$, where $L_{rod}$ is the length of the rod. 
Fig.~\ref{NNP} (bottom panel) show the master curves for 3dKA, 3dHP and 
2dMKA model (left to right). Length scale, $\xi_{rod}(T)$ is obtained by 
demanding the best data collapse. In the insets of Fig.~\ref{NNP} (bottom 
panel), $\xi_{rod}$ is plotted along with the dynamic length scale, $\xi_D$ 
obtained by other conventional methods \cite{KDSAnnualReview,KDSROPP}. 
The collapse obtained for all of 
the models are observed to be good along with the fact that the length scale 
obtained using this scaling analysis matches very well with the dynamic 
heterogeneity length scale obtained using other methods. The $\xi_D$ data
reported in this article is taken from Ref.\cite{Block}. 
This gives us the confidence that rod-like probe molecules can indeed be 
a good probe of the dynamical heterogeneity in glass forming liquids. 
Thus it is needless to mention that the proposed method can be easily
realized in experiments in the light of the already existing 
experimental results \cite{Yuan2013,Zondervan2007}. Only a systematic 
variation of the probe molecules is required to obtained the coveted 
length scale in molecular glass-formers. 

\subsection*{First Passage Time Distribution and Static Length Scale}
We now focus our attention on the decorrelation time of these rod-like 
particles immersed in a supercooled liquids with different amount of
supercooling as done experimentally in Refs.~\cite{Yuan2013,Zondervan2007}. 
In these experimental studies, it was found
that the distribution of decorrelation time is log-normal in nature 
and the width of the distribution increases with decreasing temperature.
To understand this experimental observation, we looked at the decorrelation
time of our rod-like particles in the supercooled liquid medium at
different temperatures and one indeed finds that the distribution is 
close to log-normal at least within the error bar of the simulation 
data. To gain further insight, we look at the statistics of 
``First Passage Time (FPT)'' distribution of these rod-like particles.  
First Passage distribution $F(t,\phi_c)$ in 2D is defined as the probability 
of rod crossing the angle  $\phi=\phi_c$ at a time  $t$ for the 
first time. If we unfold the $\phi$-coordinate such that 
$\phi\in(-\infty +\infty)$, then this distribution is same as that 
of well known FPT distribution of the one dimensional Brownian particle 
i.e $F(t,x_c)=\frac{x_c}{\sqrt{4\pi Dt^3}}e^{-x_c^2/4Dt}$. But the 
quantity of interest here would be the distribution of decorrelation 
time i.e $F(t,\pm \phi_c)$. This $F(t,\pm\phi_c)$ is exactly the 
distribution of time taken for a one dimensional Brownian particle 
to leave the bounded region $[-\phi_c,+\phi_c]$ while starting from $\phi=0$. 
In SI, we have shown the FPT distribution of such a Brownian rod. 
Next we compute such distribution of FPT of rod in liquid medium 
in both two dimensions(2D) and three dimensions (3D). 
%\begin{figure}[!h]
%%\vskip -0.4in
%\begin{center}
%  	\includegraphics[width=.5\textwidth , height=.4\textwidth]{Pictures/brown1.eps}
%	\caption{{Trapping time distribution of a Brownian particle in region $x\in[-x_c,+x_c]$, starting from origin. As argued in text it is same as the first passage distribution of a Brownian rod with absorbing wall at {\color{blue}$\phi=\pm x_c$ }}}
%	\label{brown}
%\end{center}
%\end{figure}

%\noindent{\bf 2D Case:}
For Brownian motion of rod in 2D, one can easily verify that
\begin{equation}
\mathcal{P}_c(\phi,t)=\frac{1}{\phi_c}\sum_{n=0}^{\infty}cos\left(\frac{(2n+1)\pi\phi}{2\phi_c}\right)e^{-\frac{(2n+1)^2\pi^2}{4\phi_c^2}Dt}
\label{FPTsoln}
\end{equation}
satisfies the diffusion equation (see Eq:SI-7) with two 
absorbing boundaries at $\phi=\pm \phi_c$. In this solution each eigenstate 
decays exponentially in time with decay rate $\frac{(2n+1)^2\pi^2}{4\phi_c^2}$, 
thus only $n=0$ eigenstate would contribute to the survival probability 
(probability that particle is not yet absorbed) at large times, implying, 
\begin{equation}
S(t)\propto e^{-\frac{D\pi^2}{4\phi_c^2}t}= e^{-t/\tau}.
\end{equation}
Thus the first passage time which can be obtained from survival 
probability via differentiation will also be exponential at long time. 
The exponential fit to the large time part of the distribution of FPT 
of the Brownian rod is found to be very good as shown in SI. 
Also one can obtain following solution for $\mathcal{P}_c(\phi,t)$ by the 
method of images,
\begin{equation}
\mathcal{P}_c(\phi,t)=\frac{1}{\sqrt{4\pi Dt}}\sum_{n=-\infty}^{\infty}(-1)^ne^{-
\frac{(\phi+2n\phi_c)^2}{4Dt}}
\label{imageSol}
\end{equation}
\begin{figure}[!htpb]
\centering
\includegraphics[width=.95\textwidth, height=0.4\textwidth]{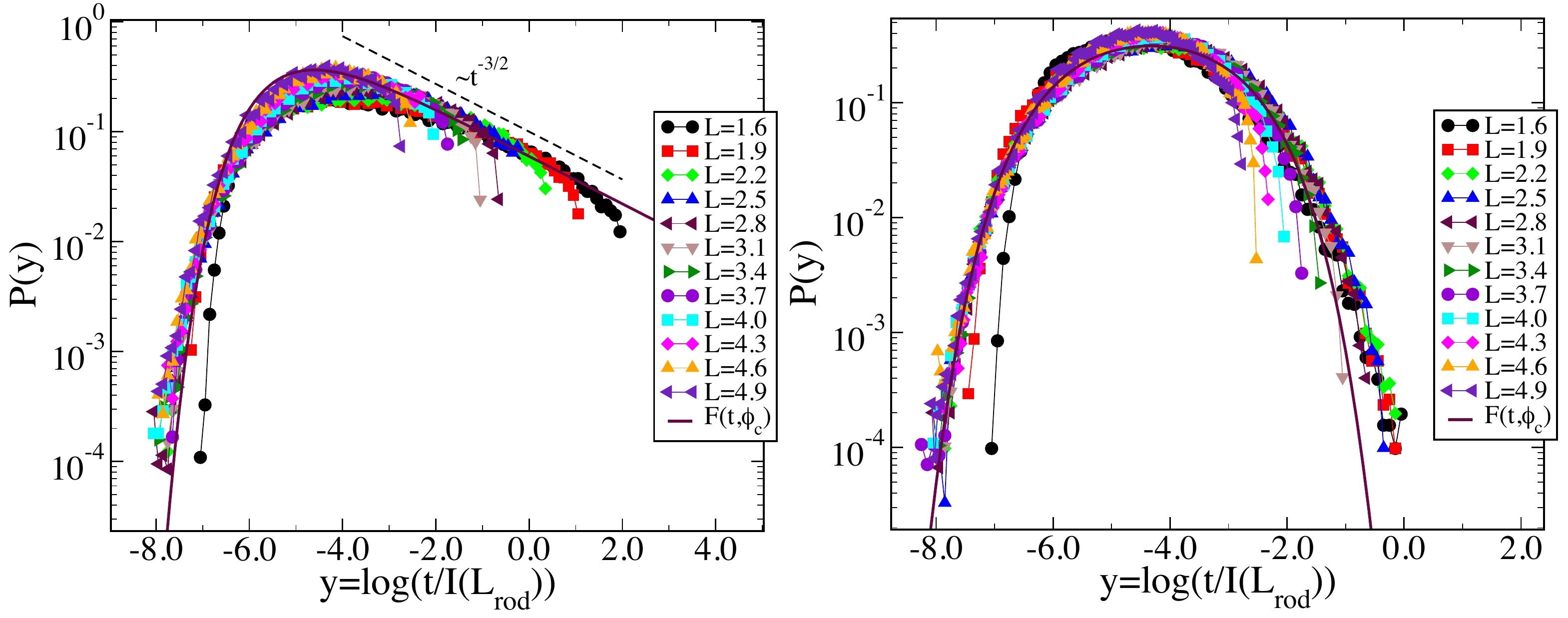}
	\caption{Left panel: Distribution of first passage times for rods of various 
	lengths scaled by moment of Inertia ($I(L_{rod})$ in 2dMKA system at 
	temperature $T=1.5$ (high temperature) with one absorbing boundary at $
	\phi=\phi_c$. Note, for this calculation we have use unbounded coordinates, i.e 
	$\phi\in [-\infty,+\infty]$.
	Right panel: Distribution of first passage times for rods of various lengths scaled 
	by moment of Inertia, $I(L_{rod})$ in 2dMKA system at temperature $T=1.5$ 
	(high temperature) with two absorbing boundaries at $\phi=\pm\phi_c$. The 
	bold line is the fitting to Eq.\ref{FPTapprox}.}
	\label{rodFPT1}
\end{figure}
Note that the two solutions, Eq.\ref{FPTsoln} and Eq.\ref{imageSol}
are same but represented via two different series. Readers are encouraged
to read Ref.\cite{Balakrishnan} for details. In the limit of small time, 
only $n=0$ term would contribute which would 
lead us to following expression of survival probability at small times.
\begin{equation}
S(t\rightarrow 0)\propto \frac{\sqrt{4\pi Dt}}{\sqrt{4\pi Dt}} erf\left({\phi_c}
{\sqrt{4Dt}}\right).
\end{equation} 
On differentiating this survival probability with a negative sign would give us 
first passage time distribution to be, 
\begin{equation}
F(t\rightarrow 0,\pm \phi_c)\propto \frac{\phi_c}{\sqrt{4\pi Dt^3}}e^{-
x_c^2/4Dt} 
\end{equation}
Thus the approximate closed form for the distribution, $F(t,\pm \phi_c)$ of rod 
and the exact series solution followed from Eq:\ref{FPTsoln} are given by the 
following expressions  
\begin{equation}
F(t,\pm\phi_c)\propto t^{-\beta}e^{-\alpha/t}  e^{-t/\tau}
\label{FPTapprox}
\end{equation}
\begin{equation}
F(t,\pm\phi_c)=\frac{\pi D}{\phi_c^2}\sum_{n=0}^\infty (-1)^n(2n+1)e^{-
\frac{(2n+1)^2\pi^2Dt}{(4\phi_c^2)}}
\label{FPTexact}
\end{equation}
With this exact solution we can obtain the mean first passage time to be $
\left\langle t \right\rangle=\phi_c^2/2D$. If we introduce a new variable 
$y=log(Dt)$ where $t$ is the first passage time then, the distribution
$P(y)$ would be independent of $D$, hence independent of rod length and 
temperature. This also implies that diffusion constant $D$ would only change 
the mean of $P(y=log(t))$ and not the shape of distribution for spatially
homogeneous diffusion as is the case in pure Brownian motion or in high
temperatures liquid state. As shown in SI 
rod length dependence of diffusion constant goes as $D(L_{rod})\sim 1/I(L_{rod}
)$ with $I(L_{rod})$ being the moment of inertia of the rod. So, similarly 
$P(y)$ with $y=log[t/I(L_{rod})]$ would be 
independent of rod length. So we can subtract out this obvious rod length 
scaling in order to understand the effect of supercooled liquid environment 
on the FPT distributions at various length scales.  
\begin{figure*}[!h]
\centering	
\includegraphics[width=.95\textwidth ]{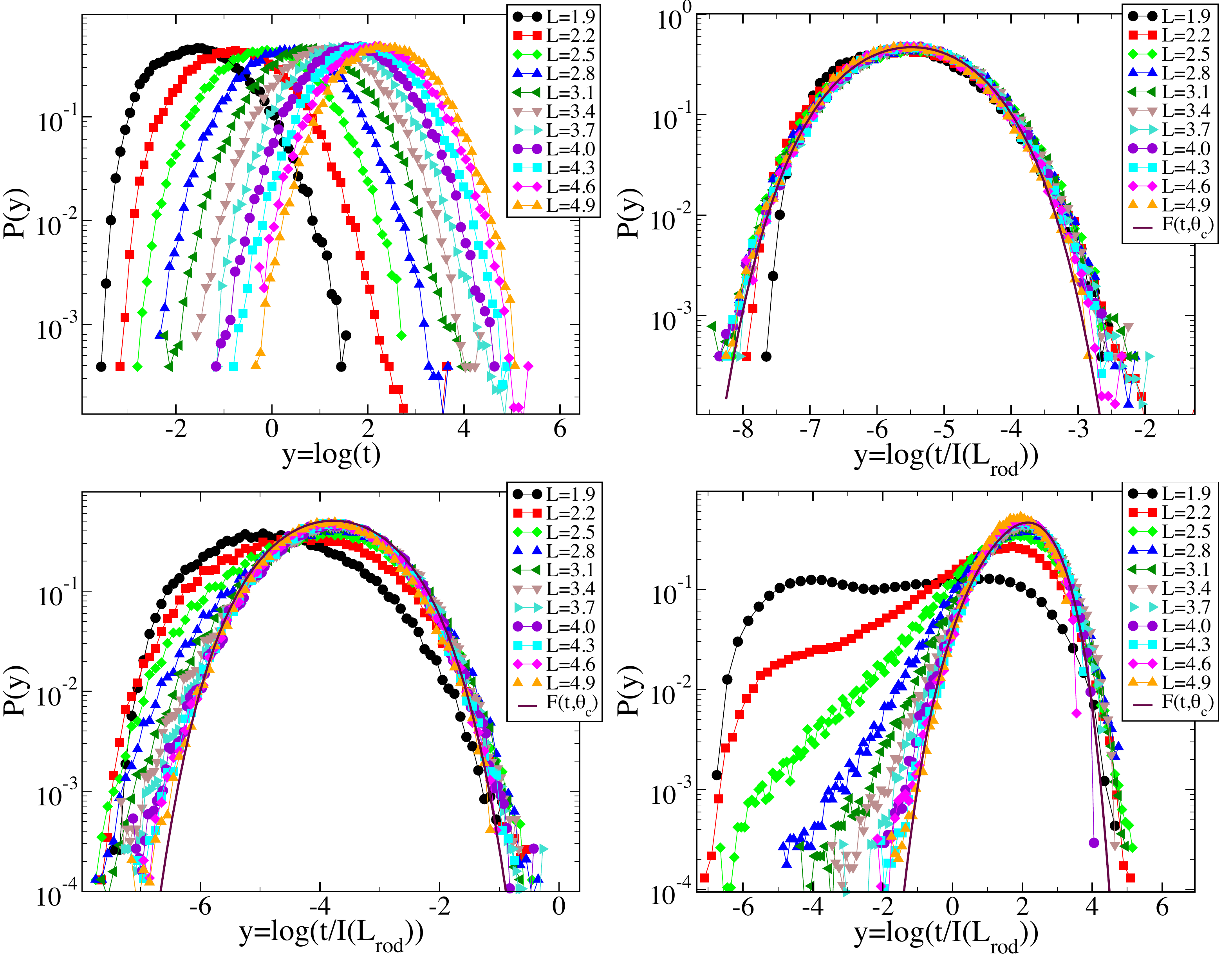}
\caption{Top panels: Distribution of unscaled (left) and scaled 
(right) first passage times (scaled with moment of inertia) for 
rods of various lengths in 3dKA system at temperature $T=2.0$ 
(high temperature) with absorbing boundary at $\theta_c=\pi/8$.
Bottom panels:	Distribution of  scaled first passage times 
(scaled with moment of inertia) for rods of various lengths in 
3dKA system at temperature $T=1.0$ (left) and $T=0.5$ (right) 
(low temperatures) with absorbing boundary at $\theta_c=\pi/8$.
Solid lines in these figures are fit to the Eq.\ref{FPTapprox}.}
\label{rodFPT3d}
\end{figure*}
\begin{figure*}[!ht]
	\includegraphics[width=1.0\textwidth ]{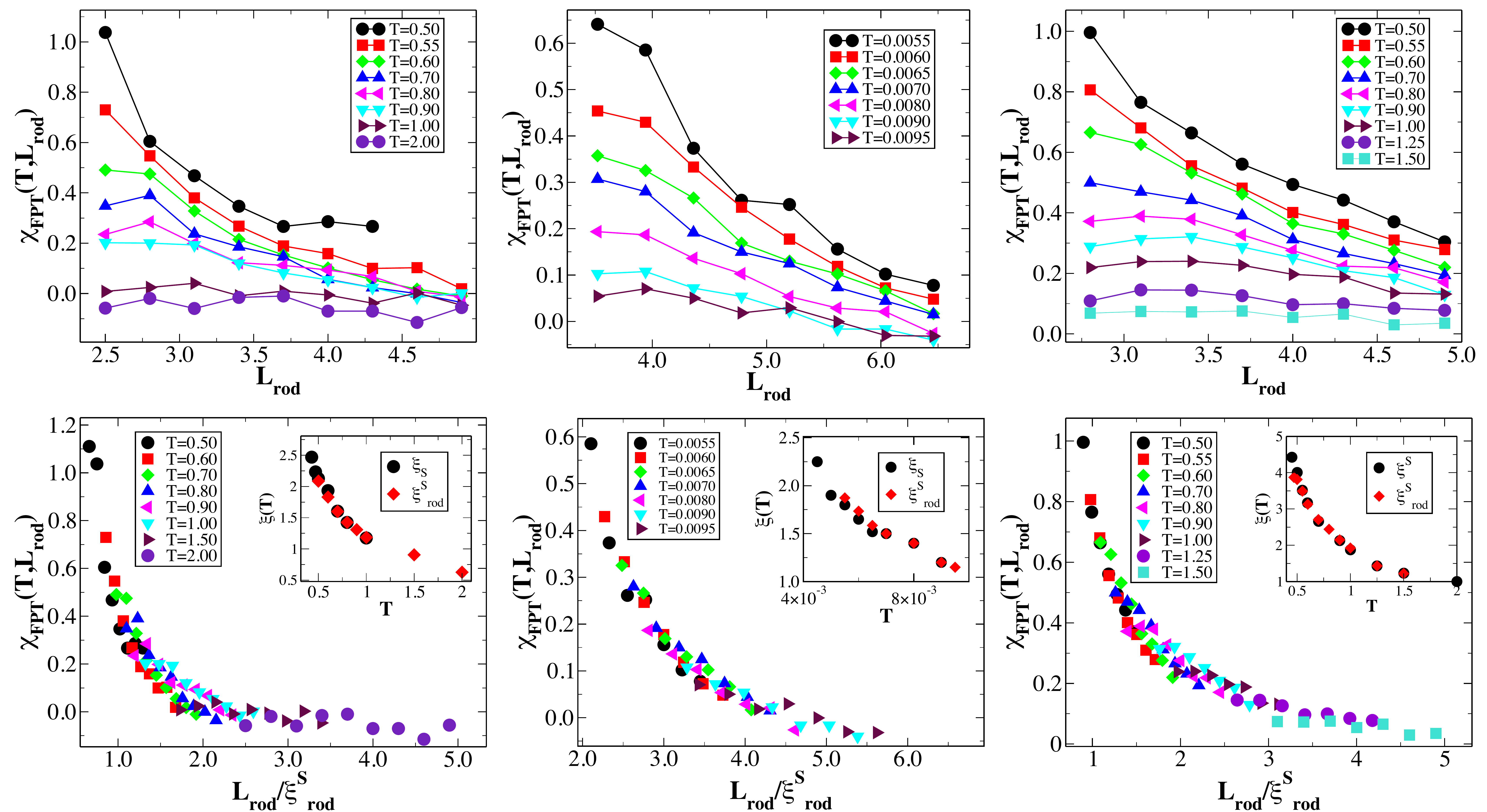}
\caption{Top panel: Variation in negative of skewness
($\chi_{FPT}$) of distribution $P(log(t))$ with rod length where $t$ is first 
passage time of rod immersed in a liquid modelled by 3dKA (left), 3dHP 
(center) and 2dMKA (right) systems at various supercooling temperatures. 
Increase in skewness (building up of shoulder in small time regime) with 
decreasing rod length and increasing supercooling can be clearly seen, which 
suggests the increase in strength of structural order at length scale of rod. 
Bottom panel: Scaling collapse of $\chi_{FPT}$ obtained by scaling rod length 
with appropriate system length scale $\xi_{rod}^S(T)$ to obtain a master curve 
for 3dKA (left), 3dHP (center) and 2dMKA (right) model systems. The 
comparison of this obtained length scale is done with static length scale $
\xi^S$ obtained with traditional methods like PTS and Finite-Size-Scaling (FSS) 
of $\tau_\alpha$ in the respective insets (data taken from  Ref:
\cite{RajuSoftMatter}).}
\label{skewCollapse}
\end{figure*}

The functional fit using Eq.\ref{FPTapprox} for the FPT distribution 
of the Brownian rod with two absorbing boundaries  and exact solution Eq:
\ref{FPTexact} is found to be very good as shown in SI and the value of 
exponent $\beta$ comes out to be $\beta=-0.5$ for 2D and $\beta=-1.0$ for 3D 
(3D case is discussed later) along with exponential decay. In Fig.~
\ref{rodFPT1}, the distributions of first passage time of rods scaled with their 
moment of inertia 
($I(L_{rod})$) are shown. Left panel shows the distribution if one considers 
only one absorbing boundary at $\phi=\phi_c$ where 
$\phi\in[-\infty,+\infty]$ and the right panel shows $F(t,\pm\phi_c)$
for two absorbing boundaries. These results are obtained by simulating 
rods of various lengths in 2dmKA system at high temperature $T=1.5$. 
Long time behaviour in Fig:\ref{rodFPT1} left panel fits very well with $t^{-3/2}$ 
as expected, while the fitting of Eq.\ref{FPTapprox} in right panel of 
Fig.~\ref{rodFPT1} is also found to be very good.

Unlike in 2D case, the first passage distribution in 3D is not 
analytically calculable. So we have solved it numerically (see SI
for details). $F(t,\theta_c)$ in 3D 
is defined as the probability that rod crosses the angle 
$\theta=cos^{-1}(\hat{u}(t)\hat{u}(0))=\theta_c$ at time $t$ for 
the first time. So basically we need to solve for density 
$\mathcal{P}_c(\theta,\phi,t)$ in the 3D rotational diffusion equation 
for a rod with absorbing cone at $\theta=\theta_c$, 
from there one gets the survival probability and eventually the 
first passage distribution (See SI for detailed discussion). 
The first passage distribution 
decays exponentially for 3D case as well, 
thus Eq~\ref{FPTapprox} will describe the first passage distribution 
of rod in 3D. Fig.~\ref{rodFPT3d} (a) \& (b) are the unscaled and 
scaled first passage distributions of rods immersed in 3dKA model 
system at temperature $T = 2.0$ (high temperature). One sees similar 
results for 3dHP model as well.

\vskip +0.1in
\noindent{\bf Effect of Supercooling on FPT distribution:}
Still now we have looked at the FPT distributions for rods in 2D as well as in 
3D systems at relatively high temperature where effect of supercooling 
can be ignored. If we now look at low temperature regime, 
we find distributions to develop shoulders at smaller time as shown in 
bottom panels of Fig.\ref{rodFPT3d}. These are the scaled distributions of 
log of first passage times for rods in 3dKA system at   $T=1.0$ and $T=0.5$
temperatures respectively. It is interesting to see that they broaden at 
small time regime for shorter rods and eventually converge to 
same asymptotic distribution for larger rods. These results can be 
rationalized if one assumes that the whole system is made of many 
domains of different mobilities as envisaged by RFOT theory as ``mosaic''
picture of supercooled liquid state. Since a smaller size rod can 
partially fit in one or two such patches it can have larger instantaneous 
torques and thus faster diffusion. On contrary, a larger rod would be 
in many such patches thus would show the bulk like homogeneous 
behaviour. Skewness of the distribution, $P(log(t),\pm\phi_c)$ can then be a 
good measure of this local order at a length scale comparable to the size 
of the probe rod. Thus by measuring the skewness of the distribution of FPT
for various lengths of the rod at different temperatures, one would be able
to extract the length scale of amorphous order by performing systematic
scaling analysis.

In top panels of Fig.\ref{skewCollapse}, we have shown the negative of 
skewness ($\chi_{FPT}$) of the distribution shown in Fig.\ref{rodFPT3d} for 
3dKA (top left), 3dHP (top middle) and 2dmKA (top right) models respectively. 
We have done the scaling collapse of the data to obtain the underlying length 
scale. We just scaled the x axis by a suitable choice of the length scale and 
plotted the dataas a function of $L_{rod}/\xi^S_{rod}(T)$. The data collapse 
obtainedis reasonably good and the corresponding length scale is plotted in 
the insets along with the static length scale obtained using other 
conventional methods like Point-to-Set (PTS) method and finite size 
scaling (FSS) of $\alpha$-relaxation time of the systems\cite{RajuSoftMatter}. 
Near perfect match of the temperature dependence of $\xi^s_{rod}$ with that 
of PTS length scale, suggests that first passage time distribution of the rod 
indeed captures the static length scale in the system.

%One might thus think that it will be even more noisy in experiments so skewness may
%not be a better measure of the statistics. Thus we have devised another
%measure which we refer as $\chi_{FPT}$ and defined as
%\begin{equation}
%\chi_{FPT}(T,L_{rod}) = \int_{-\infty}^{y_{max}}(y-y_{max})^2P(y,T,L_{rod})dy,
%\end{equation}
%where {\color{blue}$y = \log{(t/I(L))}$} is the logarithm of the scaled FPT.
%$y_{max}$ is the peak position of the distribution, $P(y,T,L_{rod})$. Note that $y$ has a trivial
%rod size scaling which increases with rod size as $N^3$, where $N$ is the 
%number of beads that constitute the rod (see SI for further details). 
%In the bottom panels of Fig.\ref{collapse1}, we have shown the scaling
%collapse of $\chi_{FPT}(T,L_{rod})$ using again a suitable choice of 
%length scale $\xi^S_{rod}$ to obtain the best data collapse. Bottom 
%Left panel shows the data for 3dKA model and bottom right panel shows
%the data for 3dHP model respectively. Insets show the comparison of
%the obtained length scale, $\xi_{rod}$ with that obtained from PTS
%method. The quality of the data for this new measure is much improved 
%and the data collapse obtained is also good. This gives us the confidence
%that the obtained length scales are quite reliable and this new measure
%will be suitable for experimental measurements. 

\begin{figure*}[!ht]
\hskip -0.1in	
\includegraphics[scale = 0.46]{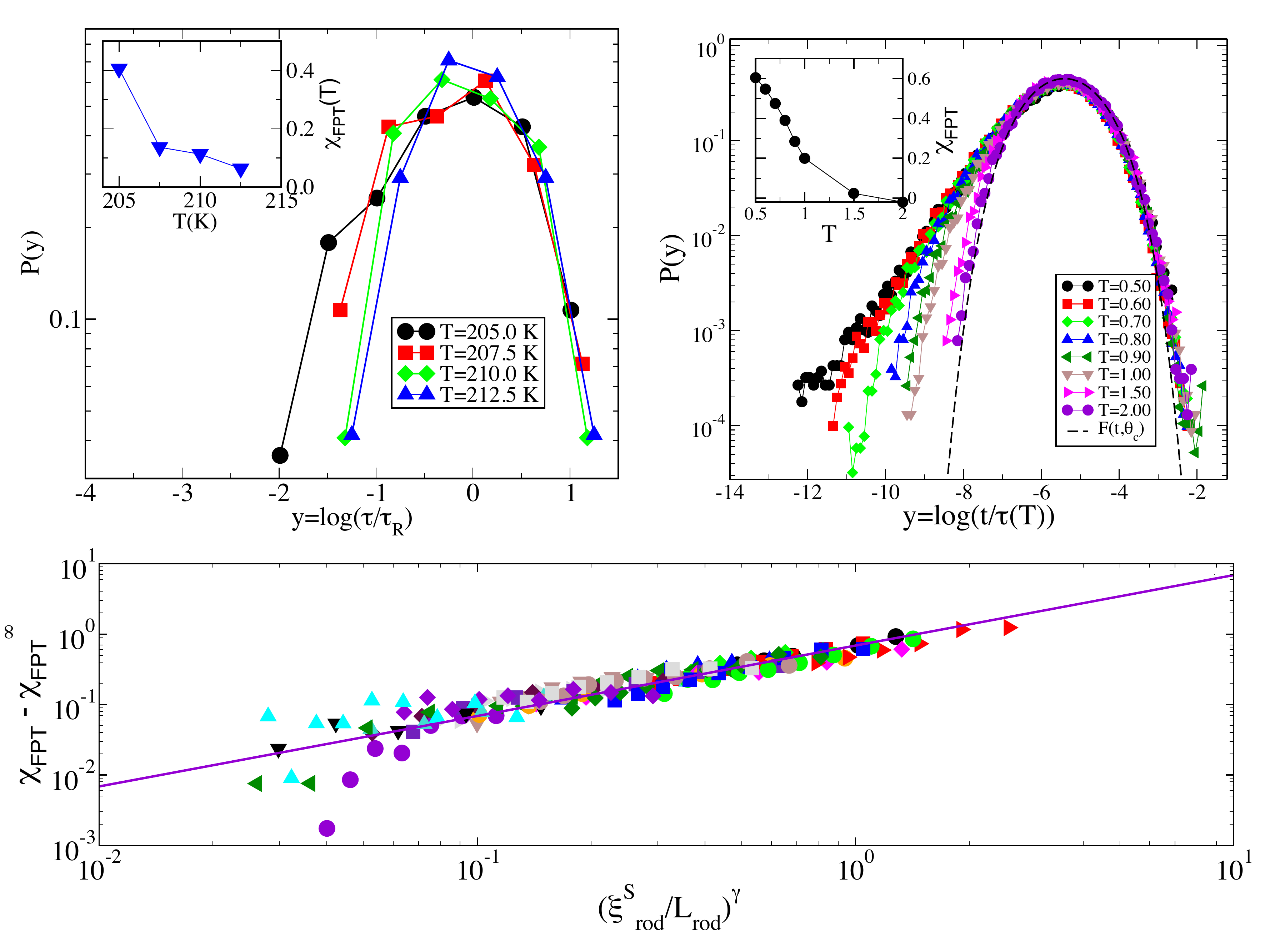}
\vskip -0.1in
\caption{Top Left panel: Distribution of correlation time of Dye molecules 
	in supercooled glycerol for four different temperatures. Data is taken
	from Ref.\cite{Zondervan2007}. The distribution is rescaled by the mean 
	correlation time. Note the clear signature of short time shoulder in the 
	distribution at lower temperatures. Inset shows the skewness of the 	
	distributions.Top Right panel: Distribution of first passage time for a rod of 
	length $2.8$embedded in 3dKA model for various temperatures. These 
	distributions alsoshow the development of shoulder (excess wing) in the 
	distribution at lower temperatures. The dashed line in through the data 
	points is the best fit to Eq.\ref{FPTapprox}. The inset shows the skewness of 
	the distribution. The similarity between experimental data and simulation 
	results are very striking. Bottom panel: Shows the skewness, $\chi_{FPT}$
	for all the three model systems plotted as $(\xi^S_{rod}/L_{rod})^{\gamma}$
	with $\gamma = 1.5$ for 3dKA and 2dmKA models and $\gamma = 3.0$ 		
	for 3dHP model respectively. The nice scaling collapse suggests that 
	$\chi_{FPT} \sim (\xi^S_{rod})^\gamma$ for a given rod length at large 
	value of $\xi^S_{rod}$. See discussion in the text.}
	\label{expt}
\end{figure*}

\vskip +0.2in
\noindent{\bf Existing Experimental Results:}
After understanding the underlying relationship between skewness of 
the distribution of first passage time or the rotational relaxation
time of the rods with the static correlation length of the 
host supercooled liquid medium, we turn our attention to reanalyze the
existing experimental results reported in Ref.\cite{Zondervan2007}. In right 
panel of Fig.\ref{expt}we show the rescaled distribution of rotation relaxation 
time, $\tau$ of the probe dye molecule in supercooled glycerol,
scaled by the time at which the peak appears.
% {\color{red}respective mean value, $\tau_R = \langle \tau \rangle$}. 
The distribution clearly shows that with increasing supercooling 
$P[\log(\tau/\tau_R)]$ starts to show shoulder as 
seen in simulation results for small or intermediate size rod length at 
low temperatures as shown in the right panel of the same figure.
In the inset of the left panel, we show the calculated skewness from 
experimental data, $\chi_{FPT}$ which seems to decrease 
very systematically with increasing temperature in complete agreement 
with our simulation results (see the inset of right panel).
This also confirms the growth of amorphous order in supercooled glycerol
as suggested in Ref.\cite{Albert2016}. In the right panel figure, we show how 
the distribution of first passage time changes with temperature for a rod 
length of $2.8$ in 3dKA model. The distributions are scaled by $\tau(T)$, the 
time at which the peak of the distribution appears at that temperature. This 
simple rescaling collapses the large time part of the distribution completely 
across different temperatures, only smaller timescale part of the distribution 
shows gradual growth of a shoulder with decreasing temperature. The inset 
shows the skewness of the distributions. The striking similarity between 
experimental data and simulation data are indeed very encouraging. If one 
does similar analysis of the experimental data of rotational correlation
time of gold nanorod in supercooled glycerol from Ref.\cite{Yuan2013}, 
then one sees very little change in the skewness of the distribution (see SI). 
This is also in good agreement with our results as it suggests that
for larger size probe molecules (gold nanorod is much larger than dye
molecule) the skewness will be smaller. 
In bottom panel, we show the dependence of skewness, $\chi_{FPT}$ 
for all the model systems at different length of the rod as well as
at different temperatures. $\chi_{FPT}$ plotted as a function of 
$\xi^S_{rod}/L_{rod}$ seems to show a power law relation at large
value of the argument. $\chi_{FPT}^\infty$ is the skewness of the 
FPT distribution for large rod lengths. It is close to zero for
almost all the temperatures.
The exponent ($\gamma$) of the power law 
turns out to be not universal across different models. So we plotted
the data as a function of $(\xi^S_{rod}/L_{rod})^\gamma$ to collapse
all the data in one master curve. The collapse and the power law
seems reasonably good, suggesting that 
$\chi_{FPT} \sim (\xi^S_{rod})^\gamma$ is probably robust across 
different systems. The value of the exponent $\gamma$ is $1.5$ for
both 3dKA and 2dmKA model and $3.0$ for 3dHP model. A detailed 
understanding of this relation and the value of the exponent 
is lacking at this moment. Thus 
if one can extract the exponent, $\gamma$ for supercooled glycerol,
then one might be able to even directly compute the growth of the 
static length scale without performing the scaling analysis. 
Nevertheless, this scaling relation does suggest that skewness of the
FPT distribution is a direct measure of static correlation in 
supercooled liquids. 
Although at this moment, we are not able to 
estimate the growth of amorphous order in supercooled glycerol due 
to lack of experimental data at different length of the probe molecules 
in supercooled glycerol, but we have clearly demonstrated the 
generality and the strength of the proposed methodology for 
measuring the growth of static correlation in experimentally 
studied glass-forming liquids.

\section*{Conclusions}
To conclude, we have studied the dynamics of rod-like particles in 
supercooled liquid medium with varying lengths of the rods. We showed that 
rotational motions of the rod start to show non-normal behaviour once the 
host medium is in supercooled regime. By analyzing the variation of 
non-normal measure with increasing rod length at timescales equal to the $
\alpha$-relaxation time of the supercooled liquid, we are able to obtain the 
dynamic heterogeneity length scale of the liquid at that instance of time. We 
then showed that the distribution of the relaxation time or the first passage 
time of the rod in our simulation studies are in complete agreement with 
experimental results obtained using gold nano-rods and other elongated rod-
like dye molecules in supercooled glycerol. Our complete statistical analysis 
of the first passage time distribution of the rod shows that the 
problem can be exactly mapped into a Brownian motion of the rod with 
two absorbing boundary conditions. The asymptotic form of the 
distribution obtained from the exact results for a Brownian particle 
with two absorbing boundary condition shows remarkable match with the 
obtained distribution of the first passage time from simulations for all 
the studied glass-forming model liquids. Our results thus also establish 
that the distribution of rotational relaxation time in experiments is 
not log-normal rather has a completely different form which can be
analytically obtained by solving the equation of motion of Brownian rod with
two absorbing boundary conditions. In the experimental works it was
claimed that increasing width of the relaxation time of the rod is a direct
indicator of the growing dynamic heterogeneity in the system, but it was not
clear how to obtained the underlying length scale from these results. Our
results suggest that to quantitatively obtain the dynamic heterogeneity length 
scale and the static length scale, one needs to extend these experimental 
studies by systematically changing the length of probe molecules.  
Thus we think that dynamics of rod-like probe molecule 
in supercooled liquid is an interesting and novel way to extract various
length scales of importance in glass-forming liquids and the method is 
clearly accessible for experimentally relevant glass forming liquids. We 
thus hope that our study will encourage experimentalists to extend their
single molecule probe experiments in supercooled liquids to extract the 
dynamical heterogeneity length and the static length in these systems. 

\section*{Method Section}
We have done NPT simulations for all of the three models studied. 
We refer these models as: {\bf 3dKA Model},
{\bf 3dHP Model} and {\bf 2dmKA Model} respectively. Details 
of these models can be found in the SI. 3dKA model is the well-known 
80:20 binary mixture of A and B type interacting via Lennard-Jones 
pairwise potential\cite{KA}. 3dHP model \cite{HP} is a bridge between the 
finite-temperature glasses and hard-sphere glasses and is usually 
studied in context of jamming. It is a 50:50 binary mixture of 
soft spheres with diameter ratio 1.4 and interacting via harmonic 
pair potential. 2dmKA model \cite{2dmKA} is a modified version of 3dKA model in 
two dimensions. It is 65:35 binary mixture. The temperature and pressure 
in the simulation is controlled by Brendsen thermostat and barostat 
~\cite{Allen}. A different thermostat does not change the results 
qualitatively. 

In all of the glass formers mentioned above we have added few (two in 
3dKA and one in 3dHP, 2dmKA ) rigid rods made up of variable number of 
spheres $N$, each separated by a fixed distance from the other by a 
distance of $0.3 \sigma_{AA}$, where $\sigma_{AA}$ is the diameter of 
the largest particle type (A) for 3dKA and 2dmKA models. For 3dHP model 
we used $0.42 \sigma_{AA}$. Each of the sphere 
in a rod have same mass and interacts via same potential as particles
in the host liquid. The rod length is defined as 
$L=0.3*(N-1)+1.0$ for 3dKA and 2dmKA models and 
$L=0.42*(N-1)+1.0$ for 3dHP model.
See SI for further details. Equation of motion for translational 
dynamics of spheres and center of mass (COM) of rods is integrated by 
usual Leap-Frog integrator. The
dynamics of rod's orientation vector ($\hat{\textbf{u}}$) is also integrated 
by the same integrator but strictly following the methods illustrated 
in \cite{Rapaport2004, Allen}.
%{\color{red}Since the mean square angular displacement 
%from $\hat{\textbf{u}}$ i.e, 
%$\langle|\hat{\textbf{u}}(t)-\hat{\textbf{u}}_0|^2\rangle$ is bounded, 
%we also calculated unbounded  angular displacements at each time as 
%$\vec{\phi}(t)=\int_0^t\vec{\omega}(t')dt'$.}

\vskip 0.8cm

\noindent \textbf{Acknowledgements:} 
We would like to thank Kabir Ramola and Bhanu Prasad Bhowmik for many useful 
discussions. We also like to acknowledge Vikash Pandey for his help in the 
initial part of the project. Authors are grateful to Satya Majumder 
for his comments and suggestions. We also thank Hajime Tanaka, Hajime Yoshino 
and Kunimasa Miyazaki for many useful discussion during the Bejing Meeting 2019.
This project is funded by intramural funds at TIFR Hyderabad from the 
Department of Atomic Energy (DAE). Support from Swarna Jayanti 
Fellowship grants DST/SJF/PSA-01/2018-19 and SB/SFJ/2019-20/05 are 
also acknowledged.  

\bibliographystyle{apsrev4-1}
\bibliography{RodSphere}
%\end{thebibliography}
\end{document}

\title{Supplementary Information:\\Dynamics of Rod like Particles in Supercooled Liquids - Probing Dynamic Heterogeneity and Amorphous Order}
\author{Anoop Mutneja}
\email{anoopm@tifrh.res.in}
\author{Smarajit Karmakar}
\email{smarajit@tifrh.res.in}
\affiliation{
Tata Institute of Fundamental Research, 
36/P, Gopanpally Village, Serilingampally Mandal,Ranga Reddy District, 
Hyderabad, 500107, Telangana, India } 
\maketitle 
\section{Definitions/Methods}
\subsection{van Hove Function}
The van Hove Function is defined as probability density of finding a particle `$i$'  in the vicinity of $\textbf{r}$ at time $t$, knowing that particle `$j$' was 
in the vicinity of origin at time $t=0$. 
\begin{equation}
G(\textbf{r},t)=\frac{1}{N}\left\langle\sum_{i=1}^N\sum_{j=1}^N\delta 
\left(\textbf{r}-(\textbf{r}_i (t)-\textbf{r}_j(0)\right)\right\rangle
\label{vanHove}
\end{equation}
The van Hove Function is usually decomposed into self part, $G_s(\textbf{r},t)
$ and distinct part, $G_d(\textbf{r},t)$. Self part is defined as the probability 
density of finding particle `$i$' in the vicinity of $\textbf{r}$ at time $t$, 
while knowing that it was in vicinity of origin at time $t=0$. 
\begin{equation}
G_s(\textbf{r},t)=\frac{1}{N}\left\langle\sum_{i=1}^N\delta \left(\textbf{r}-
(\textbf{r}_i (t)-\textbf{r}_i(0)\right)\right\rangle
\label{vanHoveSelf}
\end{equation}
Distinct part of van Hove function is defined by considering different $i$ and 
$j$ particles in Eq:\ref{vanHove}. Self part can also be thought as 
fraction of particles reaching in the vicinity of $\textbf{r}$ at time `$t$' while 
all starting form origin (displacement distribution function). 
If particles follow normal diffusive dynamics in isotropic medium then in the 
limit of large time and displacement self part of van Hove function is given 
by
\begin{equation}
G_s( \textbf{r},t)\xrightarrow{{t,r\rightarrow\infty}}\frac{1}{(4\pi Dt)^{3/2}}
exp\left(-\frac{r^2}{4Dt}\right)
\end{equation}
This expression is true for a Brownian particle in homogeneous medium at all 
times. Thus Non-Gaussianity of self part of van Hove function is a good 
measure for heterogeneity in the system at different times, where  
heterogeneity is meant by spatial variation of diffusion constant. Thus, 
non-Gaussian parameter, $\alpha_2(t)$ becomes the 
obvious choice to measure dynamic heterogeneity in a disordered system.   
\begin{equation}
\alpha_2(t)=1-	\frac{\left\langle|\textbf{r}|^4\right\rangle}{3\left\langle|
\textbf{r}|^2\right\rangle^2}
\label{binder}
\end{equation}

\subsection{Four-Point Susceptibility $\chi_4(t)$} 
Four-Point Susceptibility \cite{Dasgupta1991,Book1} is calculated as the 
variance or fluctuation of two point density correlation function 
or the overlap function $Q(t)$. Overlap function gives the measure of 
overlap between two configuration separated by time interval `$t$' and 
is defined as follows:
\begin{equation}
Q(t)=\sum_{i=1}^N\theta\left(a-|\textbf{r}_i(t)-\textbf{r}_i(0)|\right)
\end{equation}
where $\theta(x)$ is the usual step function and value of $a$ is chosen to 
ignore the decorrelation that might happen due to vibrational motion of 
particles in their cages. Thus the variance of overlap function 
($\chi_4(t)$) at time $t$ will be the measure of fluctuation in relaxation
in the system at that particular time due to dynamic heterogeneity. It
is defined as    
\begin{equation}
\chi_4(t)=\frac{1}{N}\left[\left\langle Q^2(t) \right\rangle-\left\langle Q(t) 
\right\rangle^2\right]
\end{equation}
Structural relaxation time $\tau_\alpha$ is also defined in terms of overlap 
function, it is the time when average value of overlap function reduces to $1/
e$ i.e. $\langle Q(\tau_\alpha)\rangle =1/e$. Note that the temperature 
dependence of $\tau_\alpha$ doesn't depend on the chosen value of $a$.
 
\subsection{Point-to-Set Method}
The existence of Mosaic state and a static length scale is proposed in 
RFOT theory. It is theoretically proposed that the static length scale 
can be measured using a special correlation function, termed as the 
Point-to-Set (PTS) correlation function~\cite{PTS}. The length scale
obtained form the analysis of PTS correlation function is generally 
called PTS length scale. A full theoretical consideration can be found 
in reference \cite{PTS}. PTS length scale ($\xi_{PTS}$) is calculated by 
taking a well equilibrated configuration of the system and then freezing 
all the particles outside a region $\mathcal{C}$ of linear dimension 
$\xi$ as a boundary condition. The particles inside $\mathcal{C}$ are 
then relaxed and overlap with the initial configuration is calculated. 
For small enough $\xi$ overlap should be perfect implying that the 
particles inside are restricted to the same equilibrium state. With 
increasing $\xi$ this overlap would vanish since particles inside 
would now explore more metastable states. The largest linear 
length scale $\xi=\xi_{PTS}$ to which volume of particles can be 
restricted to same metastable minimum is known as PTS length scale. 
Physically, it implies that mosaic of this linear dimension can survive in 
particular metastable state, and thus whole system can be thought of 
as divided into such mosaics of metastable states.
The length scale obtained in this way is same as static length scale 
obtained by finite size scaling of $\tau_{\alpha}$ and from minimum 
eigenvalue of Hessian matrix \cite{SKPhysA}.
            
\section{Rotational dynamics of rod like particles in liquid}
\noindent{\bf Two dimensional (2D) case:} 
Rotational dynamics of rod in 2D is described by rotational diffusion equation 
for distribution density $\rho(\phi,t)$,
\begin{equation}
\frac{\partial \rho(\phi,t)}{\partial t}=D\frac{\partial^2}{\partial \phi^2}\rho(\phi,t)
\label{diff}
\end{equation}   
 where, $\phi\in[-\pi,\pi]$ is the polar angle with which rod is inclined to x-
axis and $D$ is the rotational diffusion coefficient. With appropriate 
boundary condition and rod along x-axis as a initial condition, one can 
obtain the following solution,
\begin{equation}
\rho(\phi,t)=\frac{1}{2\pi}+\frac{1}{\pi}\sum_{m=1}^{\infty}\cos(m\phi) 
e^{-m^2Dt}.
\end{equation}  
The distribution density $\rho(\phi,t)$ is the self part of rotational van Hove 
function (Eq:\ref{vanHoveSelf}) for Brownian rods in homogeneous medium 
and with this solution the rotational mean square displacement (RMSD) and 
the non-normal parameter for rod dynamics can be calculated. Note that $
\left\langle cos\phi\right\rangle=e^{-Dt}$ and $\left\langle cos2\phi\right
\rangle=e^{-4Dt}$ so the RMSD would be,  
\begin{align}
RMSD(t)&=\left\langle|\hat{u}(t)-\hat{u}(0)|^2\right\rangle=2(1-\left
\langle cos\phi\right\rangle)\\
\label{RMSDeq}
&=2(1-e^{-Dt})
\end{align}	
where $\hat{u}(t)$ is the orientation vector of rod at time $t$. This 
expression for RMSD (Eq:\ref{RMSDeq}) can be used to fit simulation results 
to extract the value of rotational diffusion constant of rod. Next, in the spirit 
of finding the non-normal parameter for rotational distribution density of 
rods just like binder cumulant, \textit{Jain et. al.} \cite{Jain2017} shows that 
$\alpha_{rot,2D}(t)$ (Eq:\ref{2dNNP})  is zero for $P(\phi,t)$ which can be 
verified by just using the above averaged values. 
\begin{equation}
\begin{split}
\alpha_{rot,2D} = &\frac{1}{3}\frac{\left\langle|\hat{u}(t)-\hat{u}_0|^4\right
\rangle}{\left(\left\langle|\hat{u}(t)-\hat{u}_0|^2\right\rangle\right)^2}
-\frac{1}{24}\left\langle|\hat{u}(t)-\hat{u}_0|^2\right\rangle\\
& \times\left( \left\langle | \hat{u}(t)-\hat{u}_0|^2\right\rangle - 8 \right) 
-1
\end{split}
\label{2dNNP}
\end{equation}
Non-zero value of this parameter at time $t$ would signify the fluctuation 
in rotational diffusion constant at that time. Thus we have used this parameter 
in main text to quantify dynamic heterogeneity at the length scale of 
rod. 

\noindent{\bf Three dimensional (3D) case:}
The rotational diffusion equation in three dimensions reads out,
\begin{equation}
\frac{\partial \rho(\theta,\phi,t)}{\partial t}=-D\mathcal{R}^2\rho(\theta,
\phi,t)
\label{diff3d}
\end{equation}
where $\theta\in[0,\pi]$, $\phi\in[0,2\pi]$ is azimuthal angle and polar 
angle subtended by the rod and $D$ is diffusion constant. The rotational 
operator $\mathcal{R}^2$ is defined as
\begin{equation}
\mathcal{R}^2=-\left\{ \frac{1}{sin\theta}\frac{\partial}{\partial \theta}
\left(sin\theta\frac{\partial}{\partial \theta}\right)+\frac{1}{sin^2\theta}
\frac{\partial^2}{\partial\phi^2}  \right\}
\end{equation}
Solution to this equation with relevant boundary conditions can be obtained 
by introducing the spherical harmonics with the initial condition of rod 
to be along $+ve$ \textit{z}-axis, we can integrate out $\phi$ to get the 
solution,
\begin{equation}
\rho(\theta,t)=\sum_{n=0}^{\infty}\frac{2n+1}{2}P_n(cos(\theta))e^{-n(n+1)Dt}
\end{equation}
where $P_n(cos(\theta))$ is the Legendre polynomial of degree $n$. The 
average of any function $A(\theta)$ can be calculated as 
\begin{equation}
\left\langle A(\theta)\right\rangle=\int_0^\pi d\theta sin\theta A(\theta) 
\rho(\theta,t).
\end{equation}
Like in 2D, $\left\langle cos\theta\right\rangle$ and $\left\langle 
cos2\theta\right\rangle$ can be calculated easily and they are $e^{-2Dt}$ 
and $\frac{1}{3}(4e^{-6Dt}-1)$ respectively. So the rotational mean square 
displacement and the non-normal parameter\cite{Jain2017} in 3D turn out to 
be,
\begin{equation}
\left\langle|\hat{u}(t)-\hat{u}(0)|^2\right\rangle=2(1-\left\langle cos\theta
\right\rangle)=2(1-e^{-2Dt})
\label{RMSD3d}
\end{equation}
\begin{equation}
\alpha_{rot,3D} = \frac{1}{2}\frac{\left\langle|\hat{u}(t)-\hat{u}_0|^4\right
\rangle}{\left(\left\langle|\hat{u}(t)-\hat{u}_0|
^2\right\rangle\right)^2}
+\frac{1}{6}\left\langle|\hat{u}(t)-\hat{u}_0|^2\right\rangle\\ -1
\end{equation}

\section{Rotational Mean Squared Displacement(RMSD)}
\begin{figure}[!htpb]
	\centering
	\includegraphics[scale = .500]{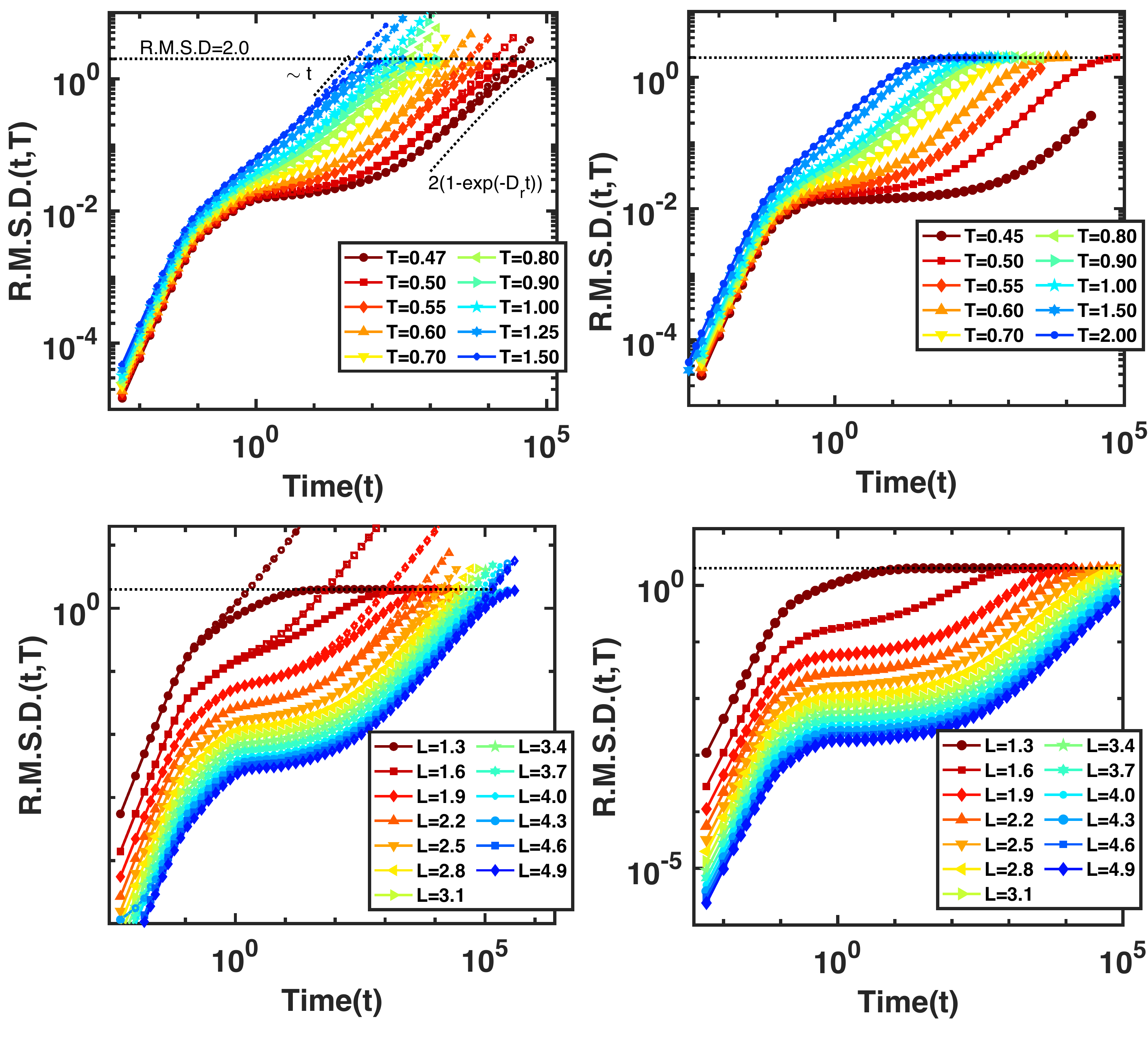}
	\caption{Top panel: Time evolution of RMSD for rod of length $L=2.5$ 
	($N=6$ beads) immersed in 2dMKA (left) and 3dKA (right) model systems at 
	various temperatures. Bottom panel: Time evolution of RMSD for rod of 
	different lengths immersed in 2dMKA (left) and 3dKA 
	(right) model systems at $T=0.5$ temperature. Dotted lines in the plots 
	2dMKA model (left) are RMSD plots in the unfolded coordinates} 
\label{RMSDFig}
\end{figure}
RMSD for orientation vector of rod ($\hat{\textbf{u}}$) is simply given by,
\begin{equation}
RMSD=\langle|\hat{\textbf{u}}(t)-\hat{\textbf{u}}_0|^2\rangle=2(1-\left
\langle\hat{\textbf{u}}(t).\hat{\textbf{u}}_0\right\rangle)
\end{equation}
where $\hat{\textbf{u}}(t)$ is the orientation vector of rod at time `$t$', $
\hat{\textbf{u}}_0$ is the initial orientation vector and angular bracket is for 
ensemble average. Fig:\ref{RMSDFig} (top panel) is the time evolution plot of 
RMSD of rod ( fixed length ) immersed in 2dMKA (left) and 3dKA (right) 
model at different temperatures. In the left panel we have also plotted the RMSD 
in unfolded coordinates $\phi\in(-\infty,\infty)$(dotted lines). One can clearly 
distinguish the ballistic, caging and diffusive regions and try to obtain rotational 
diffusion constant by fitting the diffusive region with exact expression 
Eq:\ref{RMSDeq} \& Eq:\ref{RMSD3d} for 2D and 3D model respectively. The 
complete analysis of rotational diffusion constant is done later in this document. 
Similarly, Fig:\ref{RMSDFig} (bottom panel) is RMSD evolution plot for rods of 
various lengths immersed in 2dMKA model (left) and 3dKA model (right) both 
kept at  temperature $T=0.50$. Plots for 3dHP model are similar and are not 
shown in this document.

Next we ask about the possible scaling function $B(L_{rod})$ of 
RMSD evolution plots with rod length in 
$RMSD(L_{rod},T,t)=A(T)B(L_{rod})C(t)$. It turns out that in the 
ballistic and caging regime, RMSD for rods can be scaled by inverse 
of moment of inertia of rod i.e. $B(L_{rod})=(I(L_{rod}))^{-1}$ (see Fig:
\ref{ScaledRMSD}) and in the diffusive regime, the rotational diffusion constant 
$D_r(T,L_{rod})$ also falls as $1/I(L_{rod})$ with rod length. 
Below we give details of these scaling analysis. 

\subsection{Scaling of Ballistic region}
Since ballistic region refers to small time region in which the rods 
are not undergoing any collision with other rods, implying the 
rotational displacement to be 
\begin{equation}
\Delta\theta=\omega \Delta t
\end{equation}
where $\omega$ is the angular velocity of rod at initial time. So averaged 
RMSD would be,
\begin{equation}
\left\langle\Delta\theta^2 \right\rangle =\left\langle\omega^2\right\rangle 
\Delta t^2
\end{equation}
which gives us rotational drift coefficient $D^{drift}_{rot}=\left\langle
\omega^2\right\rangle$. Invoking equipartition theorem i.e. $\left\langle
\omega^2\right\rangle = fK_bT/I$ we get,
\begin{equation}
D^{drift}_{rot}=\frac{fK_bT}{I(L_{rod})}
\end{equation}
where $f$ is the rotational degrees of freedom, one in 2D and two in 3D. 
This argument completely explains the observed scaling in the ballistic 
regime.

\subsection{Scaling of Caged region} 
\begin{figure}[!htpb]
\centering
\includegraphics[width=.5\textwidth]{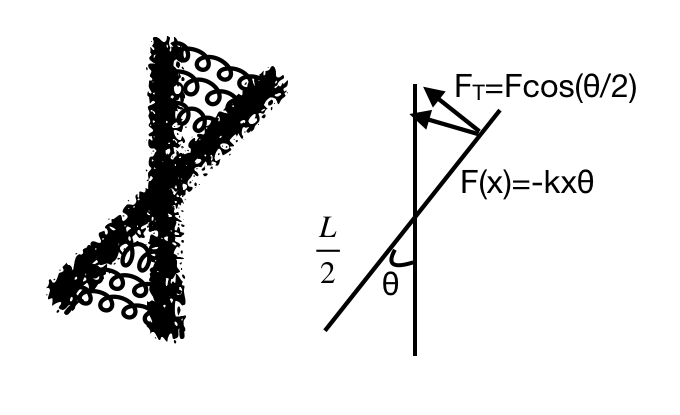}
\caption{Effective interaction picture of rod in caging regime.}
\label{caging_theory}
\end{figure}

\begin{figure}[!ht]
\centering
\includegraphics[width=.9\textwidth]{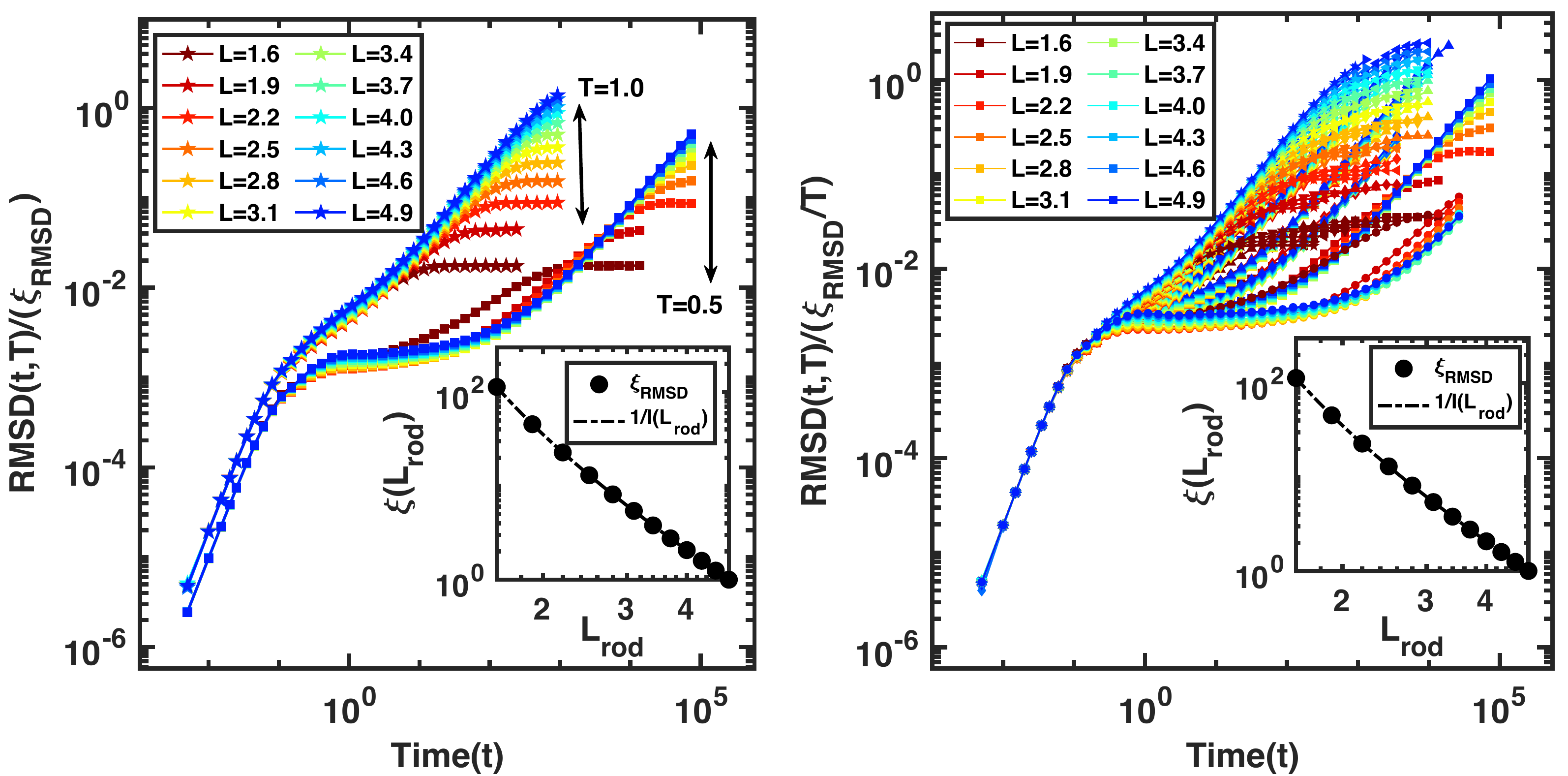}
\caption{Left panel: RMSD evolution plot scaled by $1/I(L_{rod})$  for 
rods of different length in 3dKA model liquid to get a collapse in ballistic and 
caging region. 
Right panel: Master curve of RMSD evolution plots obtained for rods of different 
lengths (different color) in systems at different temperature (different 
symbols) using a scaling function $T/I(L_{rod})$. }
\label{ScaledRMSD}
\end{figure}

Caged region refers to the time region during which the system particles 
exhibits vibrational motion. The caging RMSD value $\left\langle\Delta
\theta^2 \right\rangle _c$ is half of the vibration amplitude. Now if we 
assume each small portion ($dx$) of rod feel harmonic force with same 
spring constant $k$ from medium during this time interval, then we can 
imagine a picture like Fig:\ref{caging_theory}. In this Figure, $F_T(x)$ is the 
transverse force acting on rod element $dx$ at distance $x$ from center of 
mass of rod. If $\theta$ is the angular displacement at any time $t$, then 
transverse force on the rod element would be, $F_T(x)=-kx\theta cos
\frac{\theta}{2}$ and net torque on rod would be,

\begin{equation}
\begin{aligned}
\tau=I\ddot{\theta}&=2\int_0^{L/2} x'F_T(x')dx'\\
&=-k\left(2\int_0^{L/2} x'^2dx'\right)\theta\cos\frac{\theta}{2}\\
&=-\frac{kI}{\rho}\theta\cos\frac{\theta}{2}
\end{aligned}
\end{equation}
where $\rho$ is the mass density of rod. Since caging $\theta$ is going to 
be small thus on expanding the cosine term and keeping upto $\theta^2$ 
terms we get,
\begin{equation}
\ddot{\theta}=-\frac{k}{\rho}\theta + \mathcal{O}(\theta^3)\approx-k'\theta
\end{equation}
where $k'$ is now independent of rod length. This is the equation of 
simple harmonic motion (SHM) with 
solution $\theta(t)=\theta_0sin(\sqrt{k'}t)$ thus, $\left\langle\theta^2\right
\rangle=\left\langle\theta_0^2\right\rangle/2$.
Also, the rotational mean squared velocity would be,
$$\left\langle\dot{\theta}^2\right\rangle=k'\frac{\left\langle
\theta_0^2\right\rangle}{2}$$
and  on using equipartition theorem we get, $$\left\langle\theta^2\right
\rangle=\frac{\left\langle\dot{\theta}^2\right\rangle}{k'}\sim\frac{f K_bT}
{I(L_{rod})}$$. Since we are modeling rods in a medium at constant 
pressure so as to maintain the same particle density at different 
temperatures, the time at which RMSD changes from ballistic to caging 
region should be independent of temperature. So, it is evident that 
one can get the master curve in ballistic and caging region for all 
temperatures if we scale the RMSD with $I(L_{rod})/T$, 
Fig:\ref{ScaledRMSD} (right) is the master curve, where color code 
is for different rod lengths and different symbols are for different 
temperatures.

\subsection{Diffusive regime}
\begin{figure}[!ht]
\includegraphics[width=1\textwidth]{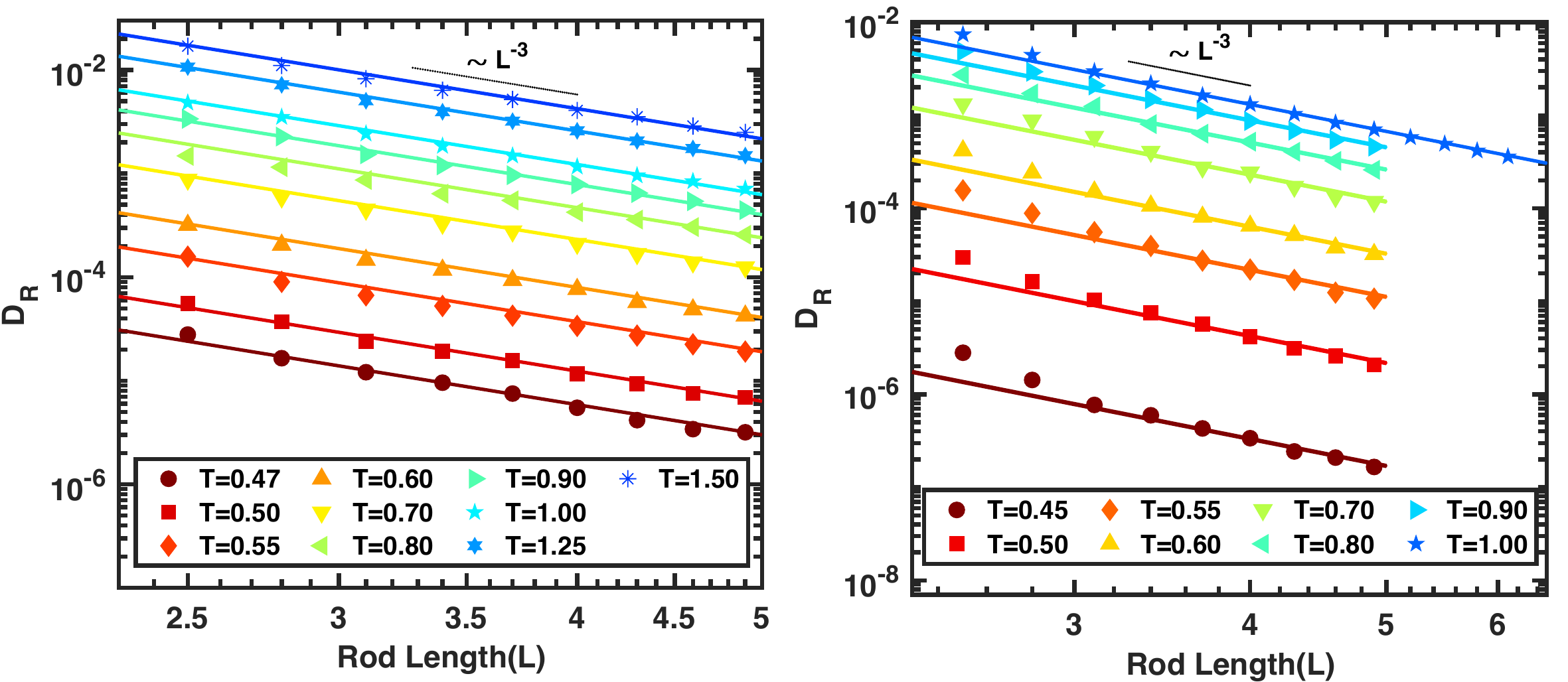}
	\caption{Rod length variation of rotational diffusion constant of rod 
	immersed in of 2DMKA system (left) and 3DKA system (right) system 
	different temperatures. Solid lines in the plot are $L^{-3}$ fit while symbols are 
	actual values of $D_R(L)$ for rod of length $L$.}
	\label{DiffCnst}
\end{figure}    
The Rotational diffusion constant can be calculated in the simulations by 
comparing the ensemble averaged value of $(\hat{\textbf{u}(t)}.\hat{\textbf{u}(0)})
$ with the exact expressions i.e. $e^{-Dt}$  for 2D and $e^{-2Dt}$ for 3D. Rod
length dependence can also be obtained theoretically by considering the Langevin 
equation of motion for rods embedded in assembly of spheres. This will be 
a valid approximation for large rods. The Langevin equation reads out,
\begin{equation}
I\frac{d^2\theta(t)}{dt^2}=-\zeta (L_{rod})\frac{d\theta(t)}{dt}+\tau(t)
\end{equation}
where $\theta = cos^{-1}(\hat{\textbf{u}}(t).\hat{\textbf{u}}_0)$, $\zeta
\dot{\theta}$ is the frictional torque and $\tau(t)$ is the rapidly fluctuating 
torque with $\left\langle\tau(t)\tau(t')\right\rangle=2\zeta K_bT\delta(t-t')$. 
For large enough rods, the usual assumptions like $\theta$ independence of $
\tau(t)$ and the large difference in the timescales of fluctuations in 
$\theta(t)$ and $\tau(t)$ are valid. So, on doing the standard calculation 
to obtain the rotational mean squared displacement one would get 
following expression for rotational diffusion constant,
\begin{equation}
D(L)=\frac{2K_bT}{\zeta(L)}
\end{equation}
On assuming the rod to be a stack of beads (Shish-Kebab model) and each of 
which would experience the viscous force according to Stokes law, one can 
get the rod length scaling of $\zeta(L_{rod})\sim I(L_{rod})~\sim L_{rod}^3$ ( page 
no 291-293 of Ref:\cite{BookPolymer} ). Hence, one gets the scaling for rotational 
diffusion constant to be $D_r\sim I(L_{rod})^{-1}\sim L_{rod}^{-3}$. Fig:
\ref{DiffCnst} shows the variation of diffusion constant of rod with changing 
rod length when they are immersed in 2dMKA (left) and 3dKA (right) supercooled 
liquid at different temperature.

Another check for the scaling of $\zeta(L_{rod})$ would be to invoke fluctuation-
dissipation theorem i.e. 
\begin{equation}
\zeta(L_{rod})=\frac{1}{K_bT}\int_0^\infty\left\langle\tau(t')\tau(t'+t)\right
\rangle dt
\end{equation}
and check for the rod length dependence of the integral. For delta correlated 
torques it is just the variance of the torque distribution on the rod. It has 
been checked and confirmed in simulations that the integral in the right side of 
above equation scales like $L^3$ for large rods. While this scaling doesn't 
hold for small rods which is not so difficult to understand because the 
Langevin approach would not work if size of Brownian particle is comparable 
to the liquid particle.
%One can also explain this scaling of rotational diffusion constant based on 
%the fact that the time needed for a large enough rod to break a cage 
%($t_{cage}$) is only dependent on temperature, because it is the time in 
%which spheres closer to the end of the rod diffuse, since they are the one 
%responsible for caging. Number of such spheres would become constant for 
%larger rods hence the cage breaking time. As caging $\left\langle
%\theta^2\right\rangle\sim 1/I(L_{rod})~\sim L^{-3}$ implies,
%\begin{equation}
%D=\frac{\left\langle\theta^2\right\rangle_c}{t_{cage}}\sim I(L_{rod})^{-1}
%\sim L^{-3}
%\end{equation}    

\section{Non-Normal parameter and dynamic length scale}
\begin{figure*}[!h]	
\includegraphics[scale = 1.150]{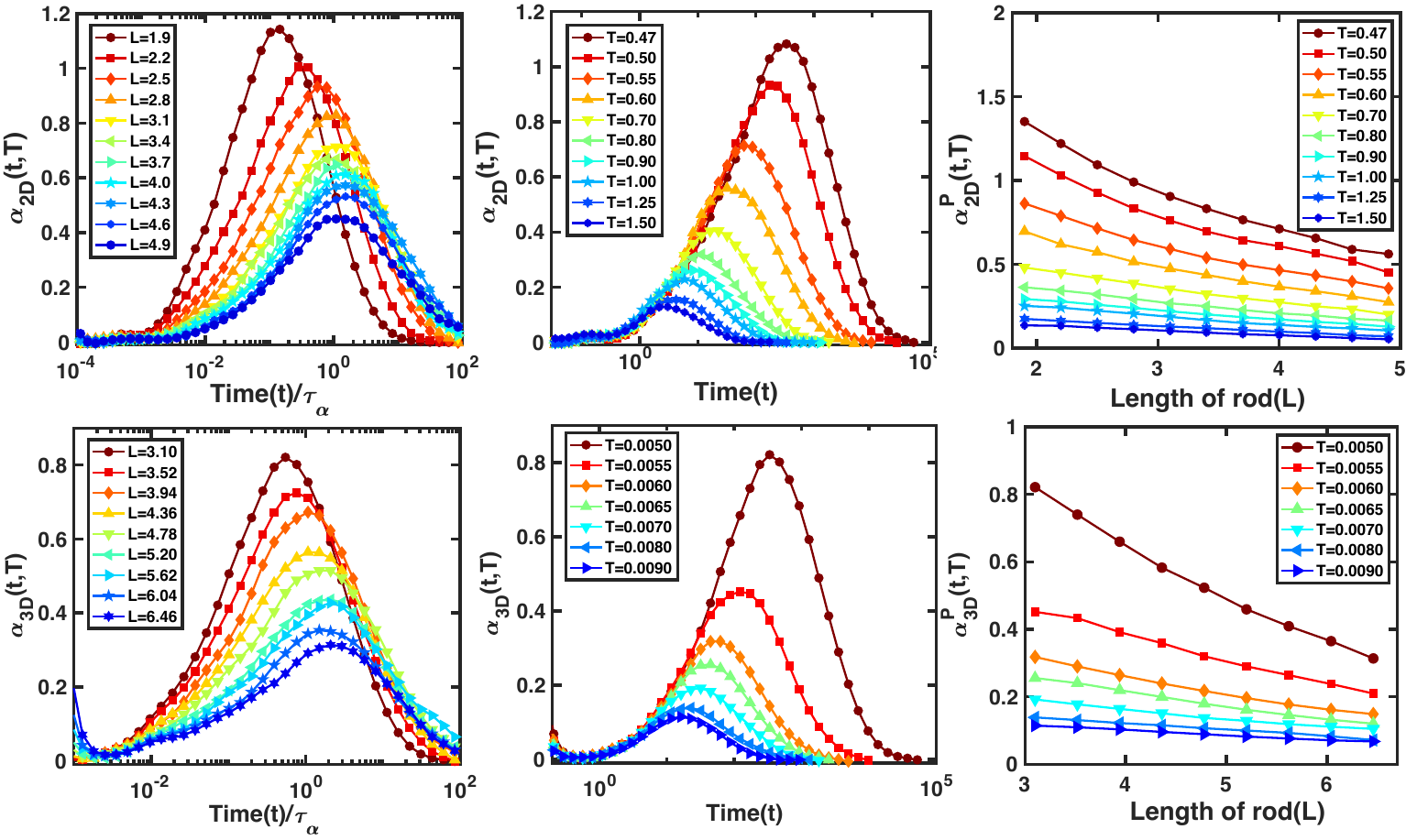}
\caption{ Left panel:  Evolution of non-normal parameter ($\alpha_{2d}$ or $
	\alpha_{3d}$) for rod of different length immersed in 2dMKA (top) 
	and 3dHP (bottom) model system at temperature $T=0.5$ and $T=0.0050$ 
	respectively. 
	Centre panel: Time evolution of $\alpha_{2d}$ or $\alpha_{3d}$ 
	for rod of length $L_{rod}=2.5$ and $L_{rod}=3.10$ in model system 2dMKA
	(top) and 3dHP (bottom) respectively.
	Right panel: Rod length variation of peak value of NNP (	$\alpha^P_{2d} $ or $	
	\alpha^P_{3d}$) for rod immersed in 2dMKA (top) and 3dHP (model) at various 
	temperatures.}
	\label{NNP2dMKA3dHP}
\end{figure*}
The Non-Normal parameter (NNP) [ $\alpha_{rot2D}(t,T,L_{rod})$ \& $
\alpha_{rot3D}(t,T,L_{rod})$ ] is a measure of fluctuation in diffusion constant 
felt by rod at time $t$, immersed in a liquid at temperature $T$. As already 
pointed out NNP should be zero for rod immersed in homogeneous medium, 
so one should expect the same for rod in high temperature ( or low density) 
liquid at all times and for all rod lengths. But in supercooled environment as 
shown in main text, this NNP would be non zero and will reach a maximum at time 
$t\sim\tau_\alpha$ similar to $\chi_4(t,T)$. Thus the peak height can be a good 
measure of dynamic heterogeneity and its increase with decreasing 
temperature confirms that. It is also shown that the peak height ($
\alpha^P_{rot2D}$ \& $\alpha^P_{rot3D}$) decreases with increasing rod 
length because of rod seeing dynamical response of the surrounding medium 
averaged over the length scale of the
rod. A length scale $\xi_{rod}(T)$ is obtained by scaling the rod length 
with an appropriate system length scale to obtain a master curve between $
\alpha^P_{rot2D}(T)$ or $\alpha^P_{rot3D}(T)$ and $L_{rod}/\xi_{rod}(T)$. 
The collapse of $\alpha^P_{rot2D}(T)$ or $\alpha^P_{rot3D}(T)$ for all three 
model systems (3dKA, 3dHP and 2dMKA) are presented in main text and the 
length scales so obtained match very well with dynamic length scales 
obtained by traditional methods. The plots of time and rod-length variation 
of $\alpha_{rod2D}(T)$ or $\alpha_{rod3D}(T)$ at different temperatures for 
3dKA model are given in main text, while Fig:\ref{NNP2dMKA3dHP} contains  
same plots for 2dMKA (top panel) and 3dHP model (bottom panel).

\section{Rotational correlation function}
\begin{figure}[!h]
\includegraphics[width=\textwidth]{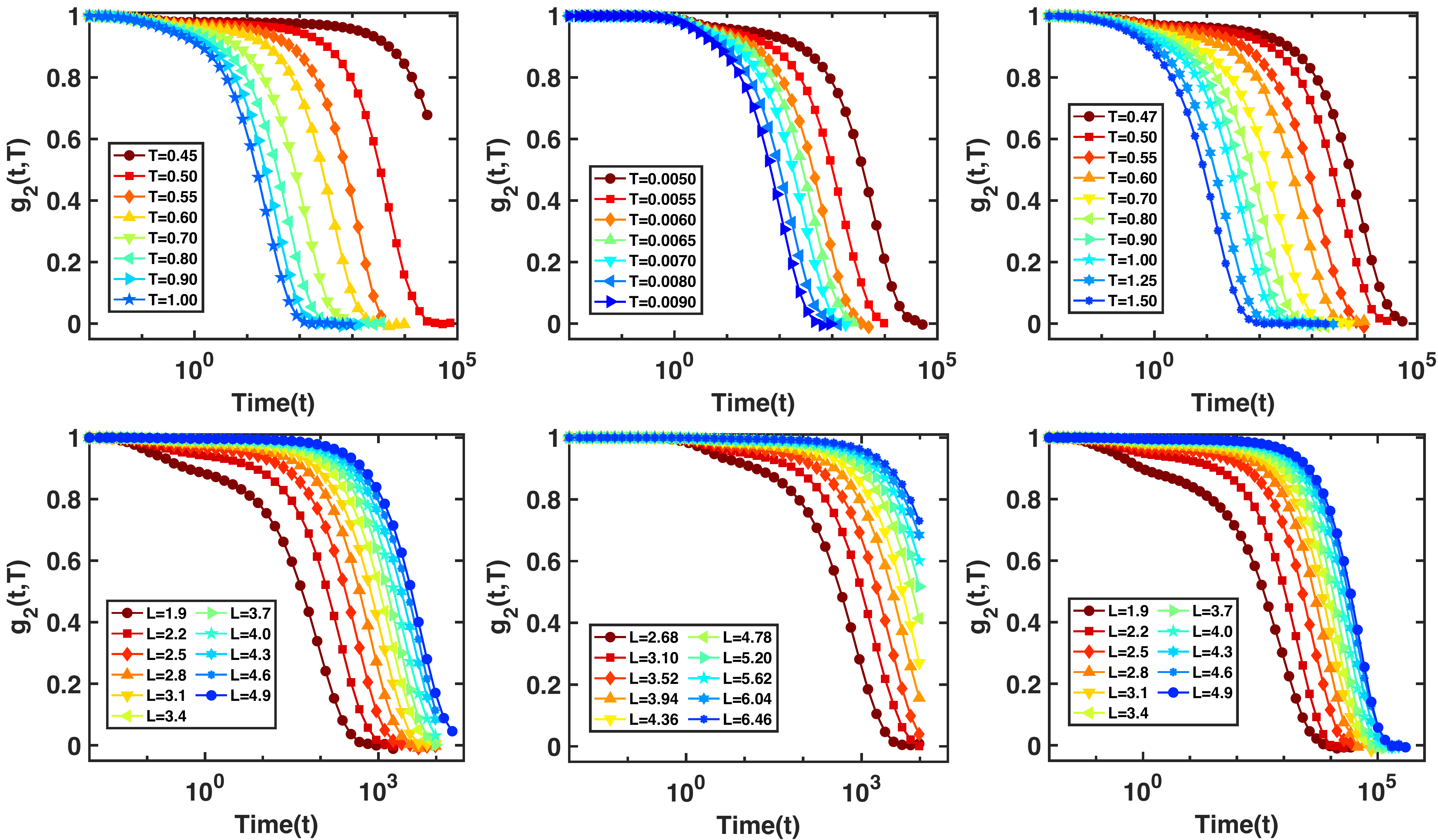}
	\caption{Top panel: Second order orientation correlation function $g_2(t)$ for rod 	
	of length $L_{rod}=2.5,3.1\&2.5$ immersed in 3dKA, 3dHP\& 2dmKA (left to right) 
	model systems respectively at different temperatures. 
	Bottom panel:  Evolution of $g_2(t)$ for rod of various rod-length immersed in 
	3dKA, 3dHP\& 2dmKA (left to right) model systems at temperatures 
	$T=0.60,0.0055\&0.50$ respectively.   }
\label{Correlation}
\end{figure}
The $l^{th}$ order orientation correlation function for 2D systems is defined 
as, 
$$g_l(t)=\left\langle cos[l\{\theta(t)-\theta(0)\}] \right\rangle$$ and for 3D 
systems it is defined as, 
$$g_l(t)=\left\langle P_l(cos[\theta(t)])P_2(cos[\theta(0)]) \right\rangle$$. 
Where $P_l$ denotes the 
Legendre polynomial of $l^{th}$ order. Both of them starts from $g_2=1$ at 
time $t=0$ and approach zero in the limit of large time. Fig:\ref{Correlation} are 
the rotational correlation functions for all three models.  Fig:\ref{Correlation} (top 
panel) shows the temperature variation of orientation correlation function of fixed 
length rod and Fig:\ref{Correlation} (bottom panel) shows its rod length 
dependence for systems at fixed (low) temperature. Two step relaxation can be 
easily seen in these correlation function indicating again the effect of 
supercooling.

\section{First Passage Time (FPT) Distribution}
As defined in the main text $F(t,\pm \phi_c)$ is distribution function for time 
when rod rotates by an angle $\phi=\pm \phi_c$ for the first time in 2D system 
while starting from $\phi=0$. In the similar way $F(t,\theta_c)$ is the distribution 
function for time when rod touches the cone $\theta=\theta_c$ for the first 
time in 3D system while starting from $\theta=0$. These can be calculated 
as a negative derivative of survival probabilities $S(t,\pm \phi_c)$ \& $S(t,
\theta_c)$, which are easy to get after solving for densities $\rho_c(t,\phi)$ and 
$\rho_c(t,\theta)$ in respective diffusion equation (Eq:\ref{diff} for 2D and Eq:
\ref{diff3d}  for 3D) with appropriate boundary and initial conditions. In the 
following sections we present the solution for FPT distributions for 2D and 3D.

\begin{figure}[!ht]
\includegraphics[width=\textwidth]{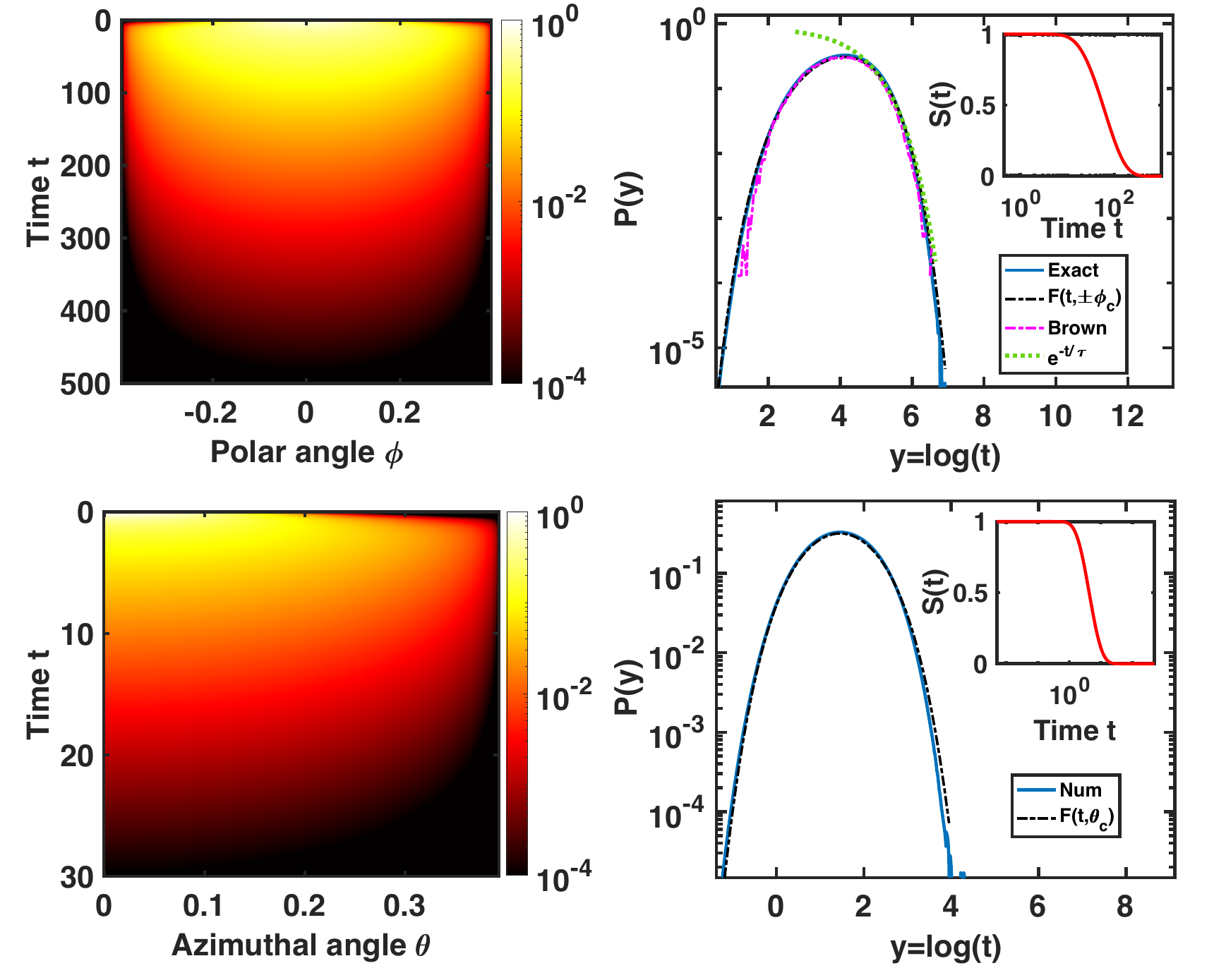}
\caption{Top panels: Heat-map (left) for $\rho_c(\phi,t)$ obtained by numerically 
integrating Eq:\ref{diff} with absorbing boundaries at $\phi=\pm\phi_c$ and initial 
condition of $\phi=0$. Right panel contains the FPT distribution $P(log(t))$ fitted 
with proposed approximate function Eq:\ref{FPTapprox}, exponential decay 
function and Brownian dynamics result along with the survival probability $S(t)$ in 
the inset. Bottom panel: Same plots as top panel but for rod in 3D. }
\label{FPTNumSoln}
\end{figure}

\vskip +0.1in 
\noindent{\bf 2D case: }
One can check that following is the solution of Eq:\ref{diff} with two absorbing 
boundaries at $\phi=\pm\phi_c$ and $\phi=0$ as initial condition.
\begin{equation}
\rho_c(\phi,t)=\frac{1}{\phi_c}\sum_{n=0}^{\infty}cos\left(\frac{(2n+1)\pi\phi}
{2\phi_c}\right)e^{-\frac{(2n+1)^2\pi^2}{4\phi_c^2}Dt}
\label{Diffsoln2d}
\end{equation}
Survival probability $S(t)$ can be calculated by integrating probability density 
$\rho_c(\phi,t)$ over whole $\phi$ space at time $t$, then a negative derivative with 
time would give the first passage distribution $F(t,\pm\phi_c)$.
\begin{equation}
S(t,\pm\phi_c)=\frac{4}{\pi}\sum_{n=0}^\infty \frac{(-1)^n}{(2n+1)}exp\left(-
\frac{(2n+1)^2\pi^2Dt}{(4\phi_c^2)}\right)
\end{equation}
\begin{equation}
F(t,\pm\phi_c)=\frac{\pi D}{\phi_c^2}\sum_{n=0}^\infty (-1)^n(2n+1)exp
\left(-\frac{(2n+1)^2\pi^2Dt}{(4\phi_c^2)}\right)
\label{FPTexact}
\end{equation}
Fig:\ref{FPTNumSoln} (top panel) shows the heat map of $\rho_c(\phi,t)$ (left) and 
the distribution $P(log(t))$ (right) where $t$ is first passage time. In the right panel 
we have also plotted $P(log(t))$ obtained from Brownian dynamics simulation as 
outlined in the main text along with the exponential decay fit. Survival 
probability is also plotted in inset of right panel. In the main text we have argued 
the following approximate functional form for these FPT distributions.
\begin{equation}
F(t,\pm\phi_c)\propto t^{-\beta}e^{-\alpha/t}  e^{-t/\tau}
\label{FPTapprox}
\end{equation}
Justification for this comes from small and large time limits of FPT 
distribution. Because of bounded space, it has to decay exponentially at large 
times (See Fig:\ref{FPTNumSoln} top-left panel) which gives us third term and at 
small times it should be like normal 
Brownian motion which would provide us first two terms (for mathematical 
arguments refer to main text). In right panel of Fig:\ref{FPTNumSoln} (top) we 
fitted above approximate function with actual solution, the fitting of function 
seems near perfect suggesting the correctness of approximate functional form 
of FPT distribution (Eq:\ref{FPTapprox}). 

\vskip +0.1in
\noindent{\bf 3D case: }
Unlike in 2D, getting solution of Eq:\ref{diff3d} with absorbing cone at $\theta=
\theta_c$ is not possible so we turn to numerical solution. In Fig:
\ref{FPTNumSoln} (bottom panel) we present the heat map (left) for $\rho_c(\theta,t)
$ and the FPT distribution (right) $P(log(t))$ with the inset of survival 
probability. The same approximate functional form Eq:\ref{FPTapprox} also works 
for very well in 3D case.

\begin{figure*}[!htb]	
\vskip +0.2in
\includegraphics[width=\textwidth	]{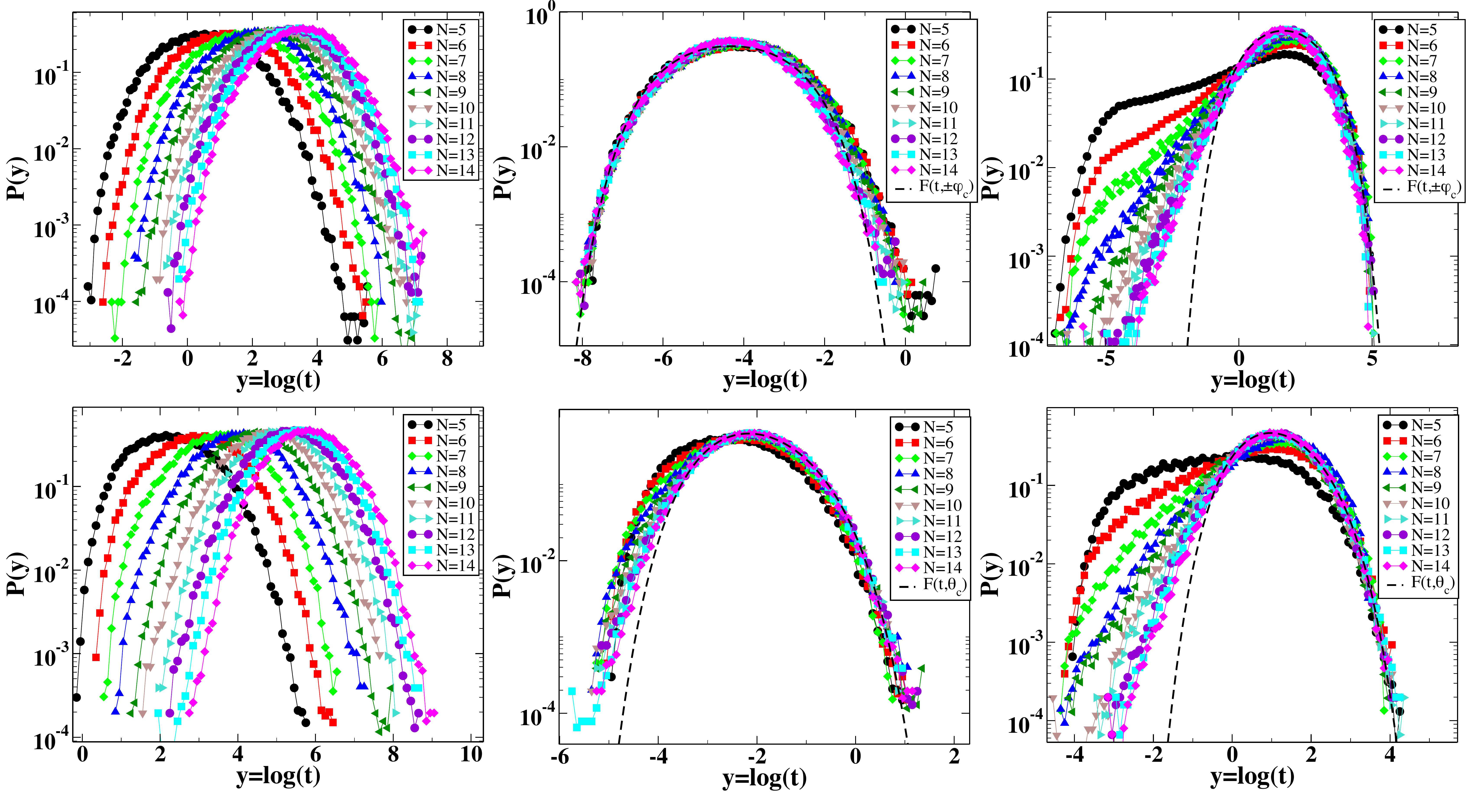}
	\caption{Top panels: Distribution of unscaled (left) and scaled (center and 
	right) first passage times for rods of various lengths in 2dMKA system at 
	temperature $T=1.5$ (high temperature) (left and center) and $T=0.5$ (low 
	temperature	) (right) with absorbing boundary at $\theta_c=\pi/8$.
	Bottom panels: Distribution of unscaled (left) and scaled (center and right) 
	first passage times for rods of various lengths in 3dHP system at temperature 
	$T=0.0090$ (high temperature) (left and center) and $T=0.0050$ (low 
	temperature	) (right) with absorbing boundary at $\theta_c=\pi/8$.}	
	\label{FPTD2DMKA3DHP}
\end{figure*}

In the main text it is mentioned that shape of the distribution $P(log(t))$, 
doesn't changes with change in diffusion constant, only mean value of the 
distribution shifts by $-log(D)$, this can be easily seen in case of 2D by writing $P(y)
$ in a following form while in 3D it has been confirmed numerically,
\begin{equation}
\begin{split}
P(y)&=\frac{\pi De^y}{\phi_c^2}\sum_{n=0}^\infty (-1)^n(2n+1)exp\left(-
\frac{(2n+1)^2\pi^2De^y}{(4\phi_c^2)}\right)\\
&=\frac{\pi e^{(y+logD)}}{\phi_c^2}\sum_{n=0}^\infty (-1)^n(2n+1)exp
\left(-\frac{(2n+1)^2\pi^2e^{(y+logD)}}{(4\phi_c^2)}\right)
\end{split}
\end{equation}
Thus, variance, skewness and all other higher order moments of distribution 
$P(log(t))$ should be independent of $D(T,L_{rod})$ for homogeneous 
diffusion and hence 
independent of rod length and temperature. This becomes the reason of 
using $P(log(t))$ to identify the structural changes (manifested in relaxation 
time) at the length scale of rod with increasing supercooling. Since diffusion 
constant $D(T,L_{rod})\propto I(L_{rod})^{-1}$ for a rod of different length 
immersed in liquid at same temperature, the distribution $P(log(t/I(L_{rod})))$ 
should be a master curve for that temperature. Thus, deviation from that 
master curve would be the measure of structural order at length scale of rod. 
In the main text Fig:3 are the distribution plots of 
$P(log(t))$ and $P(log(t/I(L_{rod})))$ for rod immersed in 3DKA model at 
different temperature. Same are plotted here for 2DMKA and 3DHP model in 
Fig:\ref{FPTD2DMKA3DHP}. One can clearly see the effect of supercooling 
appearing in the form of short time shoulder for smaller rods. Thus, 
skewness of this distribution ( $\chi_{FPT}=-skewness(P(log(t)))$ ) qualifies best to 
capture the structural order at the length scale of rod. Collapse for $\chi_{FPT}$ is 
obtained by scaling the rod length with appropriate length scale $\xi_{rod}^s$ to 
obtain master curve between $\chi_{FPT}(T,L_{rod})$ and $L_{rod}/
\xi_{rod}^s$ and is presented in main text for 3dKA, 3dHP and 2dmKA 
models. This obtained structural or static length scale matches well 
with the static length scale obtained using traditional methods like PTS 
method and finite size scaling of $\alpha$-relaxation time.
\begin{figure*}[!htpb]	
%\vskip +0.2in
\includegraphics[scale = .55	]{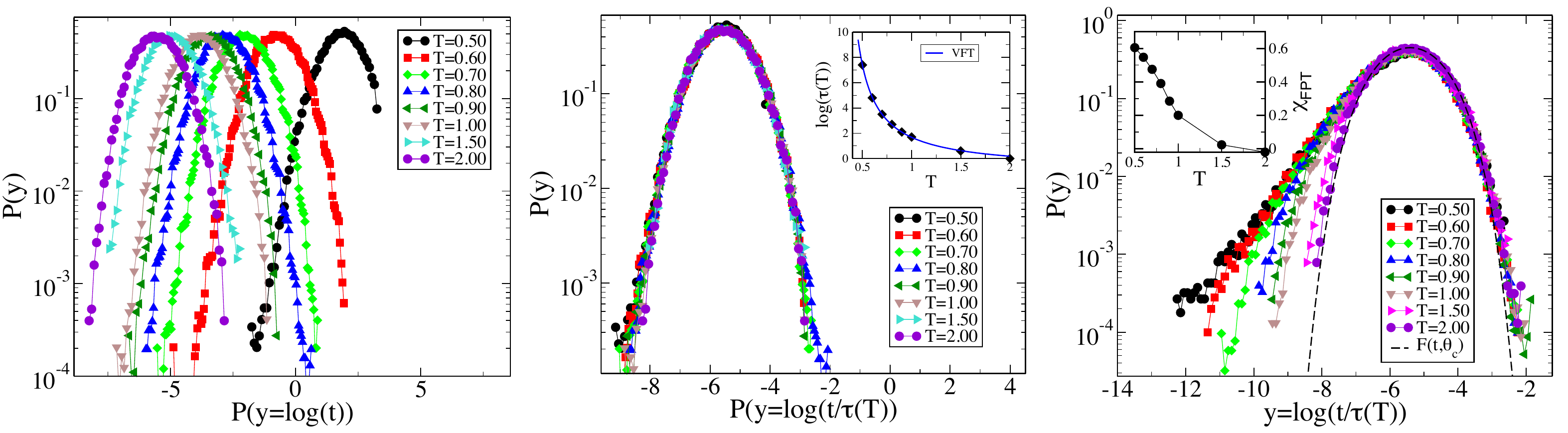}
\caption{ Left \& center panel: Distribution plots of $log(t)$ where $t$ is FPT 
	for rod length $L_{rod}=4.9$) in 3dKA system at different 
	temperatures (left) and $P(log(t/\tau(T)))$ 
	where $\tau(T)$ is chosen to obtain master curve (center). In the inset of center 
	panel $\tau(T)$ is plotted along with the Vogel-Fulcher-Tammann (VFT) fit.
	 Right panel: Distribution of $log(t/\tau(T))$ for rod of length $L_{rod}=2.8$ in 		
	 3dKA system at different temperatures, here $\tau(T)$ is same to 
	 that of $L_{rod}=4.9$ case chosen to subtract the obvious temperature scaling. In 	
	 the  inset $\chi_{FPT}$ is plotted to depict growing structural correlation at the 	
	 length scale of rod with decreasing temperature.}
\label{FPTDTempScaling}
\end{figure*}
\begin{figure*}[!htpb]
\includegraphics[scale=0.5]{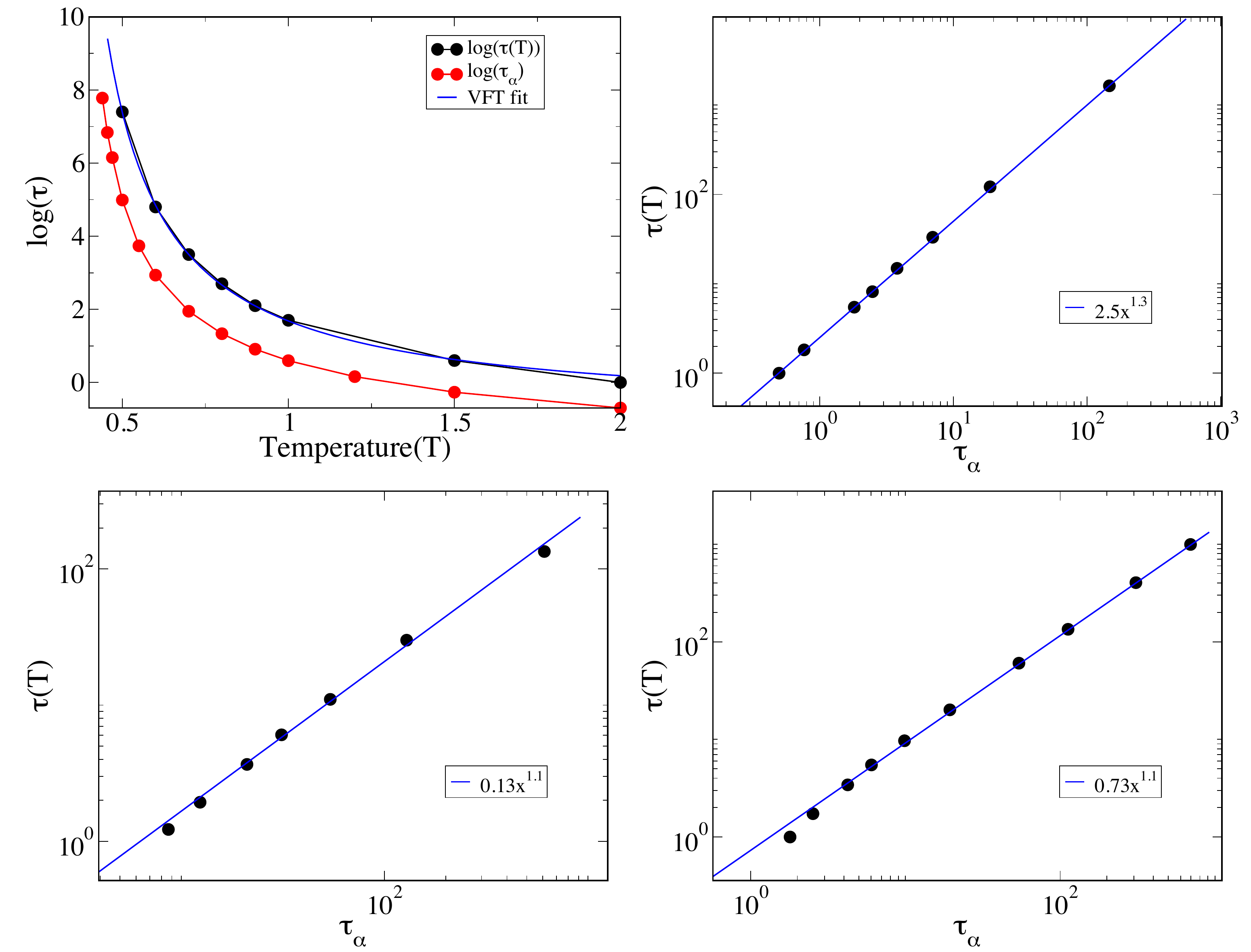}
	\caption{Top-left panel: Temperature dependence of scaling factor $\tau(T)$ fitted
	 with the Vogel-Fulcher-Tammann (VFT) law along with the temperature 	
	dependence of structural relaxation time $\tau_\alpha$. 	
	Other panels: Power law relation between $\tau(T)$ and $\tau_{\alpha}$ i.e. 
	$\tau(T)	\sim\tau_\alpha^a$, where $a=1.3$ for 3dKA model (top-right) and 
	$a=1.1$ for both 3dHP model (bottom-left) and 2dmKA model (bottom-right).  }
\label{TauRodVsTauAlpha}
\end{figure*}
Since all experiments measuring the rotational diffusion of single probe 
molecule are done for a fixed length of the probe molecules, we check 
the temperature dependence of FPT 
distribution for a fixed rod length. As argued earlier the skewness of 
distribution $P(log(t))$ shouldn't change with change in temperature for
spatially homogeneous diffusion process, but in supercooled liquid it
will show significant changes. For large enough rods this again 
shouldn't change even at low temperatures, because of averaging effect 
at larger volume. 
So one can use the temperature scaling obtained by collapsing $P(log(t))$ of 
larger rods to subtract the obvious temperature scaling from $P(log(t))$ for 
smaller rods to see the effect of growing structural order. Although it is clear 
from the value of $\chi_{FPT}$ without any rescaling. Thus increase in skewness 
($\chi_{FPT}$) of $P(log(t))$ for a rod of fixed length immersed in a liquid of 
decreasing temperature would be direct evidence of growing structural length 
scale. In Fig:\ref{FPTDTempScaling}(left \& center) the unscaled and scaled FPT 
distribution $P(log(t))$ (scaled with $\tau(T)$) for rod of length $L_{rod}=4.9$ are 
shown (3dKA model). The scaling thus obtained $\tau(T)$ is used in Fig:
\ref{FPTDTempScaling}(right) for rod of length $L_{rod}=2.8$ which collapses the 
large time part of the distributions well and extra small time shoulder is a 
measure of growing structure at the length scale of rod. The growth of $
\chi_{FPT}$ with decreasing temperature is shown in the inset of Fig:\\
\ref{FPTDTempScaling}(right) .

In Fig:\ref{TauRodVsTauAlpha}, we have 
shown the temperature dependence of $\tau(T)$. The solid line passing 
through the data points is a fit using  the Vogel-Fulcher-Tammann (VFT) 
law defined as $\tau\sim\tau_\infty exp[A/(T-T_{VFT})]$, which gives
good fit to data over a large window of relaxation time $\tau_\alpha$ 
in various supercooled liquids. The VFT divergence temperature, $T_{VFT}$ 
comes out to be $T_{VFT}^{rod} \simeq 0.273$ which is very close 
$T_{VFT} \simeq 0.296$ of the host supercooled liquid medium. Moreover
$\tau(T)$ and $\tau_\alpha$ are related to each other via power law see Fig:
\ref{TauRodVsTauAlpha} (top-right \& bottom panel), although the power with which 
both are related doesn't seems to be universal. It suggests that the
VFT divergence is actually same within error bar. Thus it is 
clear that rotational correlation time of the immersed rod can be a very
good proxy of the relaxation time of the host liquid medium.  

\section{Analysis of existing experiments}
\begin{figure}[!ht]
\centering
	\includegraphics[scale =0.5]{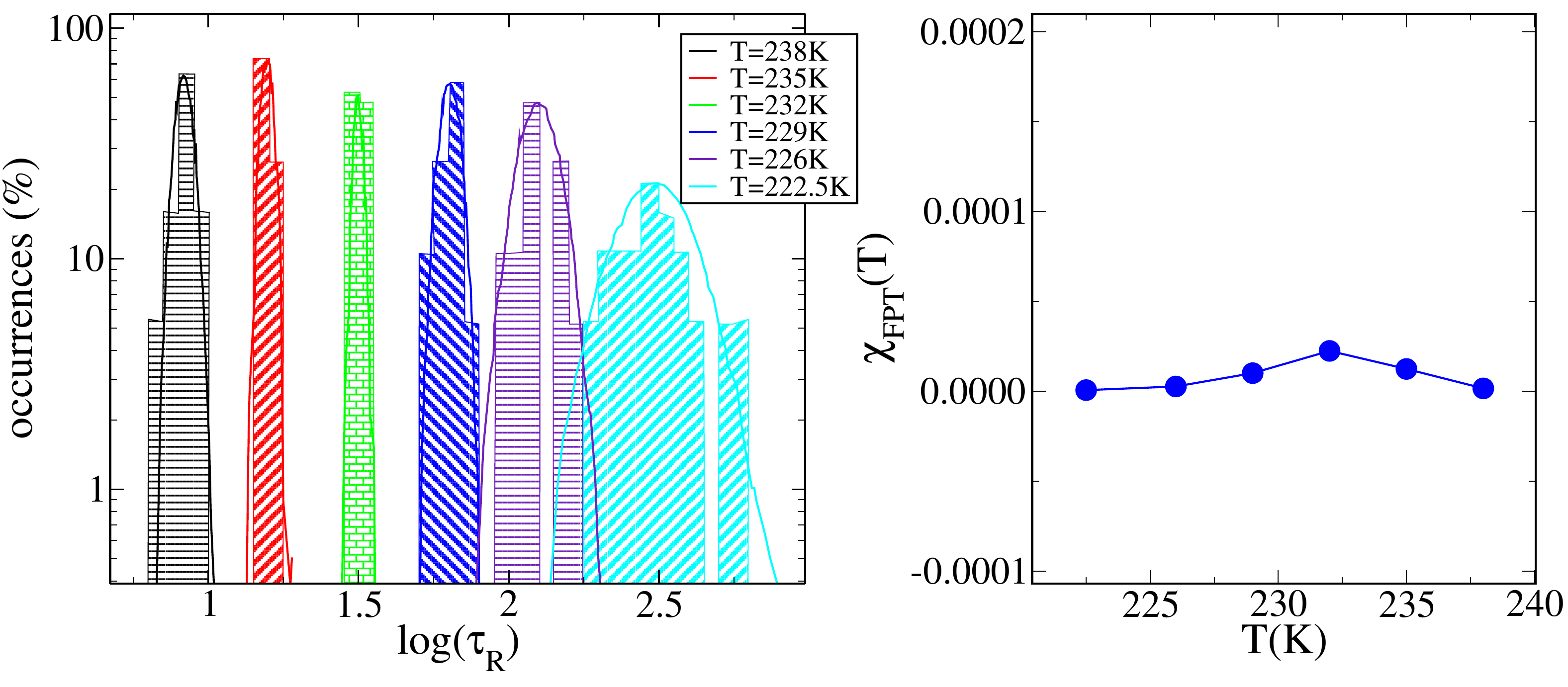}
	\caption{Left panel: Distribution of correlation time of gold nano rods immersed in 
	supercooled glycerol at different temperatures. Data taken from \cite{Yuan2013}.
	Right panel: Temperature variation of negative of skewness $\chi_{FPT}(T)$ of FPT 	
	distribution.	One can clearly see that skewness hardly changes for gold nano rods.
}
\label{ExptGold}
\end{figure} 
The existing experimental studies \cite{Zondervan2007,Yuan2013}  also uses 
the same distribution $P(y=log(\tau_R))$ where $\tau_R$ is the decorrelation 
time of the single dye molecule ($\sim 1nm$) \cite{Zondervan2007} and gold nano 
rod ($\sim30nm$) \cite{Yuan2013} immersed in supercooled 
glycerol at different temperatures. As argued earlier, if the medium is 
homogeneous (not supercooled), then the distribution should be same at 
different temperatures except the shift in mean. This is clearly not the case 
for the experiments with dye molecule, and the appearance of  shoulder 
at small time is clearly visible (see the discussion in main article).
For the experiments with nano-rod which is much larger in size than 
the dye molecules, the changes in skewness is much smaller Fig:\ref{ExptGold} in 
complete accordance with the simulation results.  
%SK this figure is not needed as we have added it in the main article.  
%\begin{figure}
%\includegraphics[width=.5\textwidth, height =.47\textwidth]{Pictures/exptDye.eps}
%\includegraphics[width=.5\textwidth]{Pictures/exptscaledDye.eps}
%\caption{Left panel: Histograms of the single-molecule rotation times 
%obtained at 210, 204, 200, and 195 K plotted on a logarithmic axis for 83 
%molecules at 210 K, 70 molecules at 204 K, 51 molecules at 200 K, and 24 
%molecules at 195 K. Solid lines are Gaussian fits to $Log(\tau_R)$. Right panel: Scaled histogram fits to match the large time decay. The appearance of shoulder in small time regime can be seen with decreasing temperature. The statistics for 195K is not much so it is asked to ignore this data set in \cite{Zondervan2007} }
%\label{ExptDye}
%\end{figure}    

\section{Model Systems}
\noindent{\bf 3dKA Model:}
3dKA ( 3d-Kob-Anderson ) model \cite{KA}is the famous 80:20 binary mixture of A 
and B type Lennard-Jones particles. The interaction potential is given by
\begin{equation}
V_{\alpha\beta}(\textbf{r})=4\epsilon_{\alpha\beta}
\left[\left(\frac{\sigma_{\alpha\beta}}{r}\right)^{12}-\left(\frac{\sigma
_{\alpha\beta}}{r}\right)^6\right]
\end{equation}
where $\alpha$ and $\beta$ varies in A, B and the interaction strengths and radii 
are $\epsilon_{AA}=1.0$, $\epsilon_{AB}=1.5$, $\epsilon_{BB}=0.5$; $
\sigma_{AA}=1.0$, $\sigma_{BB}=0.88$ and $\sigma_{AB}=0.8$. The interaction 
is truncated at $r=2.5\sigma_{\alpha\beta}$ and is made smooth by adding upto 
$4^{th}$ order terms. The simulations are done in temperature range $\
{T=0.45,1.00}$ while the pressure is so chosen that the average number density 
is, $\rho=1.2$ with 5000 particles.\\
\noindent{\bf 3dHP Model:}
3dHP ( 3d Harmonic Potential ) model \cite{HP} is a bridge between the finite-
temperature glasses and hard-sphere glasses and is usually studied in context of 
jamming. It is a 50:50 binary mixture of harmonic spheres with diameter ratio 1.4 
and interacting via potential,
\begin{equation}
V_{\alpha\beta}(\textbf{r})=\epsilon\left[1-\left(\frac{r_{\alpha\beta}}
{\sigma_{\alpha\beta}}\right)\right]^2
\end{equation}
for $r_{\alpha\beta} < \sigma_{\alpha\beta}$ and $V_{\alpha\beta}=0$ 
otherwise, where $\sigma_{\alpha\beta}=\frac{(\sigma_\alpha+\sigma_\beta)}{2}
$ and $\epsilon=1$. Again as in 3dKA, its pressure is chosen to keep average 
number density, $\rho=0.82$ with N=5000, while the temperature is varied 
between $\{T=0.0050,0.0090\}$\\
\noindent{\bf 2dMKA:}
2dMKA ( 2d Modified Kob-Anderson ) model \cite{2dmKA}is the glass forming model 
in 2 dimensions which have properties like 3dKA. It is 65:35 binary mixture of same 
A and B particles of 3dKA model interacting with exactly the same potential and 
parameters. Temperature range looked upon for this model is $\{T=0.47,1.50\}$ 
with N=1000.\\
\noindent{\bf Rods:}
In all of the glass formers mentioned above we have added few (two in 
3dKA and one in 3dHP, 2dMKA ) rigid rods made up of variable number of 
spheres $N$, each separated by a fixed distance from the other by a 
distance of $0.3 \sigma_{AA}$, where $\sigma_{AA}$ is the diameter of 
the largest particle type (A) for 3dKA and 2dmKA models. For 3dHP model 
we used $0.42 \sigma_{AA}$. Each sphere in a rod have same mass and interacts via same potential as parent spheres with $\sigma_{RodSphere}=1$, $\epsilon_{RodSphere}=1$, $\sigma_{RodRod}=1$ and $\epsilon_{RodRod}=\frac{1}{2}$ same for all models. $\epsilon_{RodRod}=\frac{1}{2}$ is chosen to avoid the chance of rods sticking together along there length. 

\bibliography{RodSphereSupli} 
\bibliographystyle{apsrev4-1}